# The gas-phase chemistry of carbon chains in dark cloud chemical models

*Jean-Christophe Loison[1,2]*, Valentine Wakelam[3,4], Kevin M. Hickson,[1,2] Astrid Bergeat[1,2] and Raphael Mereau[1,2]*

*Corresponding author: jc.loison@ism.u-bordeaux1.fr

[1] *Univ. Bordeaux, ISM, UMR 5255, F-33400 Talence, France*
[2] *CNRS, ISM, UMR 5255, F-33400 Talence, France*
[3] *Univ. Bordeaux, LAB, UMR 5804, F-33270, Floirac, France.*
[4] *CNRS, LAB, UMR 5804, F-33270, Floirac, France*

We review the reactions between carbon chain molecules and radicals, namely $C_n$, $C_nH$, $C_nH_2$, $C_{2n+1}O$, $C_nN$, $HC_{2n+1}N$, with C, N and O atoms. Rate constants and branching ratios for these processes have been re-evaluated using experimental and theoretical literature data. In total 8 new species have been introduced, 41 new reactions have been proposed and 122 rate coefficients from kida.uva.2011 (Wakelam *et al.* 2012) have been modified. We test the effect of the new rate constants and branching ratios on the predictions of gas-grain chemical models for dark cloud conditions using two different C/O elemental ratios. We show that the new rate constants produce large differences in the predicted abundances of carbon chains since the formation of long chains is less effective. The general agreement between the model predictions and observed abundances in the dark cloud TMC-1 (CP) is improved by the new network and we find that C/O ratios of 0.7 and 0.95 both produce a similar agreement for different times. The general agreement for L134N (N) is not significantly changed. The current work specifically highlights the importance of O + $C_nH$ and N + $C_nH$ reactions. As there are very few experimental or theoretical data for the rate constants of these reactions we highlight the need for experimental studies of the O + $C_nH$ and N + $C_nH$ reactions, particularly at low temperature.

## 1 Introduction

The interstellar medium (ISM) has a very rich chemistry with more than 140 molecules having been observed in various types of environments (dark clouds, star forming regions and planetary nebula). These species are formed in the gas-phase or at the surface of interstellar grains through a diverse number of processes including gas-phase bimolecular reactions, interactions with cosmic-ray particles, interactions with ultra-violet (UV) photons and chemical reactions at the surface of the grains. To study these processes, astrochemists have developed models in which the chemical composition of the gas-phase and the grain surface mantles is computed as a function of time for either fixed or time dependent physical conditions (such as temperature, density and visual extinction). The earliest models were developed forty years ago (for example (Herbst & Klemperer 1973)) and since then much progress has been made to better describe the underlying chemical processes. Modern gas-phase chemical networks contain more than 4000 chemical reactions to describe the chemistry between more than 400 species (see for instance Wakelam et al 2012). A large fraction of the reactions involved have never been studied at all or in the extreme conditions of the ISM so that their rate constants are estimated and may be assigned a very large uncertainty. Over the last few years, a significant effort has been made to identify the most important reactions in these models for future experimental or theoretical investigation. The methods used to identify important reactions are based on sensitivity analysis already described (Wakelam *et al.* 2010). Using the results of these sensitivity analyses, a number of studies have been undertaken by physico-chemists to improve the quality of the rate constants for interstellar chemical modeling (Wakelam et al. 2010, Loison *et al.* 2012, Daranlot *et al.* 2012, Wakelam *et al.* 2009). In the present study, we focus on the chemistry of carbon chain growth and adopt a different approach compared to our previous work since we consider complete reaction families rather than individual reactions.

Carbon containing molecules are ubiquitous in the interstellar medium, with hydrocarbon species containing as many as eleven carbon atoms ($HC_{11}N$) having been observed in dark clouds (Bell *et al.* 1997) and molecules such as $C_{70}$ having been observed in planetary nebula (Cami *et al.* 2010). Carbon chain growth is thought to occur mostly through C and $C^+$ reactions. The degree of hydrogenation of carbon chains is mainly driven by the reactions of $C_nH_m^+$ with $H_2$ molecules followed by Dissociative Recombination (DR). The major loss of carbon chain cations is through reaction with $H_2$ to from the corresponding protonated cation:
$C_n^+ + H_2 \rightarrow C_nH^+ + H$
$C_nH^+ + H_2 \rightarrow C_nH_2^+ + H$ and/or $C_nH_3^+ + h\nu$
Neither $C_nH_2^+$ nor $C_nH_3^+$ react with $H_2$ for $n > 2$ (McElvany *et al.* 1987, Giles *et al.* 1989, McEwan *et al.* 1999, Savic & Gerlich 2005). As a result these reactions lead to the formation of $C_nH_m^+$ with $m \leq 3$. As DR of $C_nH_m^+$ occurred through C-H bond breaking, there will be at most two hydrogen atoms on any neutral carbon chain from this pathway. In our current view of the chemistry of dense interstellar clouds, atomic radicals (C, N, O and $C^+$) remain abundant for a significant fraction of the lifetime of the cloud. Consequently, the chemistry of carbon chains (as with many other species) should be dominated by their reactions with these atoms. Whilst the elementary reactions of atomic carbon lead mainly to an increase in the size of the chains, the reactions of oxygen and nitrogen act to reduce it by breaking carbon-carbon bonds producing CO and CN radicals. In the current state of public astrochemical networks (Kinetic Database for Astrochemistry (Wakelam et al. 2012), OSU (http://www.physics.ohio-state.edu/~eric/), The UMIST Database for Astrochemistry (UdfA) (Woodall *et al.* 2007)), the estimated rate constants within certain reaction families (that is for the reactions of a specified radical species with a family of molecular coreagents which should react in a predictable way) show very different values. To improve the chemistry of carbon chains we have reviewed the reactions of the most abundant atomic radicals C, N, and O, with carbon ($C_{n=2-11}$, $C_{n=2-10}H$, $C_{n=3-10}H_2$, $C_{2n+1,n=1-4}O$, $C_{n=2-10}N$ and $HC_{2n+1}N$), proposing new rate constants for a given family based on earlier experimental measurements and the most recent experimental and theoretical studies. We then introduce these new rate constants into a dark cloud model and compare the results from the new network with earlier predictions and with observations for two C/O ratios.

This paper is organized as follows. In sections 2 (and also in the annex) we review the reactions of carbon chains with C, N and O atoms. The impact of the new reactions/rate constants for dark cloud modeling is outlined in section 3. Our conclusions are presented in Section 4.

## 2 Methodology

In this study we are interested by the reactions between carbon chain molecules and radicals, namely $C_n$, $C_nH$, $C_nH_2$, $C_{2n+1}O$, $C_nN$, $HC_{2n+1}N$, with C, N and O atoms. Most of the reactions treated in this study have unknown reaction rates (without experimental measurements and/or theoretical calculations to rely upon) in the temperature range of interest (T in the 10-200 K range). To make a reasonable estimate of their reaction rates at low temperature we use the scattered experimental and theoretical studies of similar reactions found through an extensive literature search as well as through general considerations. As the temperature is very low (near 10 K) in dense interstellar clouds, only exothermic reactions without a barrier along the reaction coordinate are important (Smith 2006, Smith 2011), with the notable exception of reactions such as F + $H_2$ (Neufeld *et al.* 2005) and possibly $C_2H$ + $H_2$ (Ju *et al.* 2005, Herbst 1994) which can occur through tunneling. To evaluate whether or not a reaction might play a role in the chemistry of dark clouds, the first step is to estimate the presence of a barrier in the entrance valley when there are no experimental measurements and/or theoretical calculations. All the reactions considered in this study proceed though addition in the temperature range 10 – 300 K. We assume that when the ground state of the adduct (complex) arises from the pairing up of electrons on the two radical reactants then there is no barrier over the Potential Energy Surface (PES), whereas if all the electrons remain unpaired then the surface is likely be repulsive (Smith 2011). So for example doublet + doublet reactions are considered to have no barrier for the singlet surface but a barrier for the triplet surfaces (which is in good agreement for H + alkyl (Monks *et al.* 1995, Harding & Klippenstein 1998) or alkyl + alkyl (Georgievskii *et al.* 2007, Klippenstein *et al.* 2006) reactions). Carbon atom reactivity is a special case as carbon atoms react with unsaturated closed shell molecules. As a result, we base our estimates on the widely available experimental and theoretical studies of carbon atom reactions.

**Rate constant estimation:**

For a given barrierless family of reactions, we first calculate the rate constant (k(T)) using capture rate theory (which leads to an upper limit value of the rate constant (Georgievskii & Klippenstein 2005)), $k(T) = g_{el}(T)\times k_{capture}(T)$, where $g_{el}(T)$ is the electronic degeneracy factor and $k_{capture}(T)$ is the rate constant given by capture rate theory dominated by long-range forces, mainly through dispersion interactions (Georgievskii & Klippenstein 2005, Clary *et al.* 1994) ( $k_{capture}(T) = 8.56\sqrt{\frac{1}{\mu}}C_6^{1/3}(k_B T)^{1/6}$ with $C_6 = \frac{3}{2}(\frac{IE_1.IE_2}{IE_1 + IE_2})\alpha_1\alpha_2$ , $\mu$ is the reduced mass, IE the Ionization Energy of each reagent and $\alpha$ is the polarisability). IE values were taken from (Lias *et al.* 1988) and polarisabilities were taken from (Woon & Herbst 2009). When no data were available, IE and polarisabilities were calculated at the Density Functional Theory (DFT) level with the hybrid M06-2X functional developed by (Zhao & Truhlar 2008), which is well suited for calculations involving main-group thermochemistry, associated with aug-cc-pVTZ basis set (Dunning Jr. 1989) using Gaussian09 (Table 2). $g_{el}(T)$ is estimated by comparing the degeneracy of the barrierless potential energy surface connecting the reagents to the products given in Table 1 (Table 1 is provided in supplementary information), with the total degeneracy of all surfaces leading from the reagents. We then compare the theoretical capture rate constants with experimental determination of some reactions of the family (for example N + CH (Brownsword et al. 1996) for the N + $C_nH$ family) or very similar reactions (C + $C_2H_2$ (Haider & Husain 1993b, Chastaing *et al.* 1999, Chastaing *et al.* 2001) for C + $C_2$, $C_3$, $C_5$, $C_7$, $C_9$). The recommended value is between the capture rate limiting one and the experimental value of a similar reaction (see Annex). For a given family the polarisability increases with the size of the reagent but this variation is offset by the variations of the ionization energy and the reduced mass, leading to very small theoretical variation of the rate constant with the size of the coreagent. As a result, we consider that the size of the coreagent has no effect on the rate constant for all reactions studied in this work. The temperature dependence of the rate constant is taken to be equal to the convolution of the $T^{0.17}$ dispersion dependence with the variation of electronic degeneracy factor, except when direct measurements of the temperature dependence exist.

In the case of carbon atom reactions, experimental room temperature rate constants for the reactions of carbon atoms with a series of alkenes, alkynes, dienes and diynes (Haider & Husain 1992, Haider & Husain 1993a, Haider & Husain 1993b, Husain & Ioannou 1997, Husain & Kirsch 1971b, Husain & Kirsch 1971a, Clary et al. 1994) up to C + 1-hexadecene and C + 1-dodecyne seem to indicate a correlation between the size of the molecule and the rate constant, with the rate constant increasing for larger molecules in contrast with the theoretical capture rate constants for the reactions of carbon atoms with alkenes which show little variation. Theoretical study shows that of the three triplet electronic states correlating with $C(^3P)$ + $^1C_2H_2$ reagents, only one presents no barrier (Takahashi & Yamashita 1996). Considering the fine structure of the carbon atom leads to an electronic degeneracy factor equal to (1+2×exp(-23.6/T))/(1+3×exp(-23.6/T)+5×exp(-62.4/T)), in relatively good agreement with the experimentally determined temperature dependence between 10 and 300 K (Chastaing et al. 1999, Chastaing et al. 2001). Then the variation of the room temperature rate constant with the size of the coreagent may indicate that either a systematic experimental error is involved or there is an increase of the electronic degeneracy factor with the size of the co-reagent. The systematic experimental error of Husain and co-worker may come from the use of a predefined mixture of coreagent and atomic carbon precursor molecule ($C_3O_2$) leading to unusual kinetic conditions without the possibility to decouple C atom reactions with the coreagent from C atom reactions with the precursor. Such experiments make it extremely difficult to determine potential secondary sources of C atoms. A comparison with other experimental measurements gives good agreement for C + $C_2H_2$ and C + $C_2H_4$ (Haider & Husain 1993b, Chastaing et al. 1999, Chastaing et al. 2001, Bergeat & Loison 2001) but not for all other reactions for which the measurements of Husain and coworkers are systematically larger, between 2.3 times and 3.0 times larger for C + allene (Husain & Ioannou 1997, Chastaing et al. 2001, Chastaing *et al.* 2000, Loison & Bergeat 2004), 1.4 times larger for C + methylacetylene (Haider & Husain 1992, Chastaing et al. 2001, Chastaing et al. 2000, Loison & Bergeat 2004), 1.5 times larger for C + propene and 1.7 times larger for C + trans-butene (Haider & Husain 1993a, Loison & Bergeat 2004). These differences are incompatible with the reported uncertainties. However, even if there is no systematic experimental error but rather an increase of the electronic degeneracy factor at room temperature with the size of the co-reagent, that will not change the value of the electronic degeneracy factor at 10 K, which will always remain close to 1. Then, the rate constants for reactions between C and unsaturated molecules should all have similar rate constant values of approximately $3.0\times10^{-10}$ cm$^3$ molecule$^{-1}$ s$^{-1}$ at 10 K. This is supported by experimental measurements for the $C(^3P)$ + $^1C_2H_2$, $^1C_2H_4$, methylacetylene and allene reactions. (Chastaing et al. 1999, Chastaing et al. 2001). So for carbon atom reactions we also consider that the size of the coreagent has no effect on the rate constant, in a similar manner to all other reactions studied in this work. This assumption is in contrast to the values currently used in astrochemical databases for C + $C_nH_2$ reactions.

Branching ratios are estimated from thermodynamic considerations, the most exothermic products having the larger branching ratio as indicated by statistical theory in the absence of exit transition states (Galland *et al.* 2003, Galland *et al.* 2001, Chabot *et al.* 2010). To determine the thermochemistry we use (Baulch *et al.* 2005) as well as the database http://webbook.nist.gov. If the enthalpies of formation were unknown we performed DFT calculations with the hybrid M06-2X functional developed by (Zhao & Truhlar 2008), which is well suited for calculations involving main-group thermochemistry, associated with cc-pVTZ basis set using Gaussian09. For the branching ratios of C + $C_n$ reactions, we use the results of Chabot e*t al.* 2010.

In total, 171 reactions have been studied and an overview of the reactivity of each family is developed in detail in the Annex. Among them, 41 new reactions (and 8 new species) are proposed and 122 rate constants have been modified for existing reactions. The reactions and proposed rate constants are listed in Table 1. In addition to the reactants and the products of each reaction, the three parameters (alpha, beta, gamma) used to derive the temperature dependence of the rate coefficients ($k(T) = \alpha\times (T/300)^{\beta}\times\exp(-\gamma/T)$) are given. For each rate coefficient, the temperature dependent uncertainty is also given: $F_0$ is the uncertainty parameter (the value of k falls within the interval between its nominal value divided by $F_0$ and its nominal value multiplied by $F_0$ for lognormal distributed uncertainties) and g is the temperature dependence parameter of the uncertainty factor ($F(T) = F_0\times\exp(g\,|1/T - 1/T_0|)$ (see Wakelam et al 2012 for more details). Uncertainty factors are difficult to estimate. When the reaction has been studied experimentally at low temperature we choose the uncertainty factor given in the corresponding publications. When the rate constant has been studied only at room temperature, or when the rate constant is deduced from similar reactions which have been well studied (particularly at low temperature) or when there are theoretical studies of the reaction, we choose an uncertainty factor of 2. When the rate constant is deduced from similar reactions which themselves are not well known (studied at room temperature only for example) the uncertainty factor is taken to be equal to 3. The g factor is in general taken to be equal to 0 as the temperature dependence of the rate constant is relatively low for barrierless reactions.

## 3 Implications for chemical models of dark clouds

### 3.1 The chemical model

To estimate the impact of these new proposed reactions and rate constants on the predicted abundances in dark clouds, we have used the latest version of the Nautilus chemical model (Hersant *et al.* 2009, Semenov *et al.* 2010). This model computes the evolution of chemical abundances for a given set of physical and chemical parameters. It takes into account the gas-phase reactions, interactions between species in the gas-phase and grain surfaces and chemical reactions at the surface of the grains. The gas-phase network is based on the public network kida.uva.2011 (http://kida.obs.u-bordeaux1.fr/models (Wakelam et al. 2012)) with a few additions to account for some important species on surfaces. The gas-grain interactions are included to mimic the sticking of gas-phase species to the surface of the grains during an encounter, the evaporation of the species from the grain surfaces due to the temperature, the indirect evaporation due to cosmic-ray grain heating (Hasegawa *et al.* 1992) and the evaporation of the products of exothermic surface reactions (Garrod *et al.* 2007). The chemical reactions at the surface of the grains are treated with the rate equation approximation (Hasegawa et al. 1992). Species at the surface of the grains are also dissociated directly by the UV photons and by the photons induced by the Prasad & Tarafdar mechanism (Prasad & Trafdar 1983). The surface network and parameters are similar to Garrod *et al.* 2007. The final list of reactions is composed of 7957 reactions for 684 species.

The chemical composition of both the gas-phase and the grains surfaces is computed as a function of time for given dense cloud physical parameters, namely a gas and dust temperature of 10 K, a H total density of $2\times10^4$ $cm^{-3}$, a cosmic-ray ionization rate of $1.3\times10^{-17}$ $s^{-1}$ and a visual extinction of 10. All elements are assumed to be initially in the atomic form, with abundances listed in Table 1 of Hincelin *et al.* 2011. Considering the uncertainty in the C/O elemental ratio and its potential importance for carbon chemistry, we have considered two different ratios: 0.7 and 0.95. In those ratios, the oxygen elemental abundance is changed whereas the carbon abundance stays at $1.7\times10^{-4}$ with respect to the total H density. As in many previous works, molecular hydrogen is assumed to be already formed at the beginning of the computation so that the initial abundance of $H_2$ is 0.5 compared to total H. The model was run for $10^7$ yr using the two C/O elemental ratios and the chemical network previously described. In addition, we updated the gas-phase network according to the propositions made in the Appendix of this paper and compared the chemical abundances computed with this updated network to the previous one. In total, we have run four different simulations: two different C/O elemental ratios were used (0.7 and 0.95) with our standard and updated gas-phase network.

### 3.2 Model results: sensitivity to the new rate constants and the C/O elemental ratio

Figure 1 shows the results for the four models for a selection of species, whose abundances have been observationally determined in two well-studied dark clouds TMC-1 (CP peak) and L134N (North peak). A compilation of observed abundances for C-bearing species is listed in Table 3. For some of the molecules, such as $C_4H$ in TMC-1, several data are available in the literature. The differences for $C_4H$ are up to a factor of seven. It is not easy to understand the reason for these disagreements and to know which data to choose. On the figure, some observed abundances are superimposed on the modeled ones. In this section, we describe the impact of the new rate constants on the predicted abundances and in section 3.3, we make some comparisons with the observations.
The effect of both the C/O elemental ratio and the new rate constants are not the same for all species and at all times. For most C-rich species, the predicted abundances are favored by the C/O elemental ratio of 0.95 and the older rate constants. In other words, the global effect of the new rate constants is to make the formation of carbon chains less efficient. One exception is $C_3N$, which is produced with larger abundances by the models using the new rate constants. All models predict similar abundances after a few $10^6$ yr, a time after which surface reactions are particularly important.

#### 3.2.1 CH and $C_2H$

The abundances of these two species are not very sensitive to the parameters (new rate coefficients and C/O ratio, see Fig. 1) but it should be noted that the new rate constants for the CH and $C_2H$ reactions with C, N and O atoms are not very different to the previous ones. The CH and $C_2H$ abundances vary by less than one order of magnitude between $10^2$ and $10^7$ yr.

#### 3.2.2 $C_nH$ (n=3-8) molecules

The predicted abundances of the $C_nH$ species are strongly reduced by the new rate constants (see Fig. 1), having a double peak profile as a function of time, with the exception of $C_6H$ and $C_8H$, whose second peak is replaced by a hole at a few $10^5$ yr for a C/O ratio = 0.7. The first peak around $10^3$ yr is due to ionic reactions, whereas the second larger peak around $(1 - 2)\times10^5$ yr is due to neutral reactions. For most C-chain species the general trends and abundances are relatively well described for both C/O = 0.7 and C/O = 0.95 (Fig. 1), the predicted abundances are larger when there is less oxygen present (with the C/O elemental ratio close to 1).

#### 3.2.3 $C_n$ (n=4-8) molecules

There are very few observations of linear carbon chains in dark molecular clouds as they are not detectable by emission microwave spectroscopy, due to the lack of a permanent dipole moment. $C_2$ molecular emission has been observed in TMC-1 (Hobbs *et al.* 1983), but is ambiguous. Indeed there are no microwave transitions for $C_2$ due to the lack of a dipole moment. As a result, $C_2$ is detected through its electronic transitions in the visible region. However, photons in the visible region are absorbed by dust particles within the molecular cloud so that $C_2$ is only observed at the edge of the cloud (a region with a density estimated equal to $10^3$ cm$^{-3}$) and does not correspond to the same location as the species detected in the microwave region with a density estimated equal to $2\times10^4$ cm$^{-3}$ for which the abundances are calculated in the model. There are notable changes with the new rates, changes which are proportional to the size of the skeleton. In particular, the simulated $C_n$ abundances are smaller with the new rates because these species react more quickly with oxygen atoms to form eventually $C_3$, which accumulates because of its low estimated reactivity.

### 3.2.4 $C_{n=3\text{-}8}H_2$ molecules

There are important changes using the new rates, with $C_nH_2$ molecules being produced more slowly, particularly for n>3, resulting in lower abundances at $2\times10^5$ yr. Moreover, the abundances decrease with the increasing size of the carbon backbone as shown in Fig. 1. It should be noted that $C_nH_2$ (n = 3-8) molecules were the most abundant family of carbon chains in simulations employing the old network. Simulations using the new network give similar abundances for $C_nH_2$ species to those obtained for other carbons chains $C_nH$, $C_n$ and $HC_nN$. There are only two molecules in this family that have been observed in dark clouds: l-$C_3H_2$ and c-$C_3H_2$ (see Table 3). Note that the observations of $C_4H_2$ and $C_6H_2$, sometimes used for comparison, concern the cumulene isomer ($H_2CCCC$), which is not the isomer included in current networks. This family is very likely to be important and is undoubtedly not well described in the present network, particularly as only one isomer is considered, except for $C_3H_2$, and the reactivity is very different between isomers.

### 3.2.5 $C_nN$ species

The sensitivity of the CN abundance to the C/O elemental ratio is different for simulations using the old and new networks, although the general effect is not very strong (see Fig. 1). CN abundances increase by only a factor of two using the new reaction rates with the model using a C/O ratio = 0.7 and this value increases to more than an order of magnitude for the model with C/O = 0.95. Before $10^5$ yr, the CN molecule is easily destroyed by reactions with atomic N and O and its chemical lifetime is about $10^3$ yr. After $10^5$ yr, there are many important CN production pathways, both neutral and ionic, the proportion of each being highly dependent on various parameters such as the C/O elemental ratio. The main CN production pathways around (1-3)$\times10^5$ yr are N + $C_2N$, O + $C_2N$, $HCNH^+$ + $e^-$, N + $C_2$, N + CH, N + c-$C_3H$, N + l-$C_3H$, and N + $C_4$. The decrease of the CN abundance around $2\times10^5$ yr for a C/O = 0.7 arises because oxygen atoms deplete later than nitrogen atoms so that CN production through N atom reactions decreases much more rapidly than CN loss through the CN + O reaction. If the observed CN relative abundance is really as low as a few $10^{-10}$ (see Table 3), this may indicate that the atomic oxygen abundance is still high at the time of the observation and then rather in favor of low C/O ratio. Alternatively, this may indicate that the rate constants for the N + $C_n$ and N + $C_nH$ reactions (as the main direct or indirect sources of CN) are overestimated at low temperature.

Despite the fact that we have changed significantly the chemistry of $C_nN$ (n = 2 - 7) species, the effect on the predicted abundances is only moderate. The early peak in the simulated abundances of $C_3N$, $C_5N$ and $C_7N$ around $10^3$ yr is due to ionic reactions.

### 3.2.6 $HC_nN$ (n=3-9) species

The abundance of cyanopolyynes is predicted to increase strongly with time with the new rates as the predicted large decrease of $HC_nN$ abundances between $10^3$ and $10^5$ yr is due to the newly introduced C + $HC_nN$ reactions (Li et al. 2006) which become important mechanisms for $HC_nN$ loss. As shown in Fig. 1, the $HC_3N$ and $HC_5N$ abundances show a peak in abundance shifted towards later times by the new rate constants, in good agreement with observations for TMC-1 with an age of a few $10^5$ yr. The $HC_5N$ abundance depends more on the C/O elemental ratio whereas $HC_7N$ and $HC_9N$ strongly depend on both the C/O ratio and the new rate constants up to $2\times10^6$ yr. $HC_7N$ and $HC_9N$ are less abundant with a C/O ratio = 0.7 and this effect is accentuated using the new rates except at very late times. Nevertheless, the chemical network for these long chain cyanopolyynes is very simplified considering in general only one isomer for large species (not only for cyanopolyynes but also for all radicals, molecules and ions involved in the formation of long chains) and the large predicted decreases of $HC_7N$ and $HC_9N$ in the 0.2 - $1\times10^6$ yr range for C/O = 0.7 may not be real.

### 3.2.7 Negatively charged species

Anions show a double peak profile in their abundances (see Fig. 1) with a strong peak early on ((1-2)$\times10^3$ yr when the electron density is still high) and a later peak after a few $10^5$ yr (when carbon chains are abundant). The intensity of the second peak depends on both the C/O ratio and the network used. A small C/O ratio coupled with the new network produces the smallest abundance of anions. One exception is $C_3N^-$, whose abundance is slightly increased by the new network for times between $2\times10^5$ and $1\times10^6$ yr, as for $C_3N$. Note that anion formation is thought to occur through radiative association between radicals and electrons; processes which are only poorly understood (Herbst & Osamura 2008) and are highly dependent on the precision of the calculations as well as on the uncertainties on simulated radical abundances. This may explain the surprising over-production in the models of $C_4H^-$ for TMC-1 considering the relative under-production of $C_4H$ as $C_4H^-$ is mainly produced in the models through the $C_4H + e^- \rightarrow C_4H^-$ radiative association reaction.

### 3.2.8 CO, OH, NO and $O_2$

The new proposed rate constants have also an impact on the abundance of other species such as oxygen containing compounds (see Fig. 1). NO and OH abundances are not particularly sensitive to the C/O elemental ratio. The change in their abundances with the new network are largely due to the current update of the rate constants for the N + OH and N + NO reactions according to new measurements and calculations (Daranlot *et al.* 2011, Bergeat *et al.* 2009). The NO abundance is directly related to the concentration of OH since it is almost entirely produced by the N + OH reaction. NO is mostly destroyed by the N + NO and C + NO reactions. The CO abundance is not much changed. The sensitivity of $O_2$ to the new rate constants depends on the C/O elemental ratio (Hincelin et al. 2011). By using a C/O elemental ratio close to 1 and the model employing the new network allows us to reduce the peak abundance of $O_2$ down to the observational limit in dark clouds. In the model with the higher C/O ratio, the oxygen atom concentration is lower after $10^5$ yr compared to the model with C/O = 0.7, leading to a lower $H_3O^+$ abundance and then a lower OH abundance (the $H_3O^+ + e^-$ dissociative recombination reaction is by far the main OH source in dark and cold regions where photochemistry does not play any role (Hollenbach *et al.* 2009)). Since $O_2$ is mostly formed by the reaction between O and OH, its abundance is reduced accordingly.

### 3.2.9 $C_{2n+1}O$ species

With the exception of $C_3O$, these species are newly introduced. The chemical network for the $C_{2n+1}O$ species is relatively small. For $C_3O$ the two main pathways of production are:

$C + C_3H_3^+ \rightarrow C_4H_2^+ + H$ followed by
$C_4H_2^+ + O \rightarrow HC_4O^+ + H$ and then
$HC_4O^+ + e^- \rightarrow CH + C_3O$ (1a)
$\rightarrow C_3H + CO$ (1b)
and
$O + c,l\text{-}C_3H \rightarrow CH + C_3O$ (2a)
$\rightarrow C_3H + CO$ (2b)
the main destruction pathways of $C_3O$ are its reaction with atomic carbon as well as its reactions with $H_3O^+$, $H_3^+$ and $HCO^+$. It should be noted that the $^3C_4 + {}^3O$ reaction is assumed to produce only $^1CO + {}^1C_3$. As a result, the abundance of $C_3O$ depends directly on the branching ratios between $CH + C_3O$ formation and $C_3H + CO$ formation through reactions (1) and (2). In the model, the pathways leading to $C_3O$ formation have been estimated to be equal to 0.5 for the DR reaction (1) and 0.3 for the $O + c,l\text{-}C_3H$ reaction. Then the overestimation of the $C_3O$ abundance (see Fig. 1) may be due to an overestimation of this branching ratio or related to the over production of $l,c\text{-}C_3H$ in the model versus the observations. Theoretical calculations of these branching ratios are clearly needed for better simulations of $C_nO$ abundances.

**3.2.10 Results for the model without gas-grain interactions**

Before $10^5$ yr, the differences between the gas-phase abundances computed using models with and without gas-grain interactions are small. After that time, the depletion of gas-phase species onto grain surfaces (and the possible reinjection of species formed on the grain surfaces into the gas-phase) has an impact on the gas-phase abundances. Only considering the CO abundance observed in TMC-1 (CP) and L134N (N), the chemical model (with the old or new networks and with C/O ratios equal 0.7 or 0.95) predicts an age of $2\times10^4 - 7\times10^5$ yr for TMC-1 (CP) and $5\times10^3 - 1\times10^6$ yr for L134N (N). It may then be necessary to consider the effect of grains for the chemistry in these two clouds. The effect of the C/O elemental ratio on computed abundances is weaker using the gas-grain model than a pure gas-phase model (Hincelin et al. 2011). This is due to the fact that large quantities of carbon (in the case of the C/O ratio close to unity) are stored on the grain surfaces and are transformed into large carbons chain molecules that remain on the grains (Garrod et al. 2007).

## 3.3 Comparison with observations

### 3.3.1 Individual species

The diversity of observed values for the abundances of molecules published in the literature, very often without error bars, makes the comparison between model predictions and observations difficult for some molecules. In addition, it must be mentioned that the uniform approximation of the cloud structure, used both for our chemical modeling and the published observations (within the beam size and along the line of sight), increase the difficulty of making such comparisons.

For most species observed in TMC-1 (CP), we can find an agreement for a specific time and C/O elemental ratio with one of the published observed values but sometimes for a cloud age outside the reasonable age range (a few $10^5$ yr) given by the CO abundance. Exceptions are $C_4H$, which is under-produced in the simulations and $c,l\text{-}C_3H_2$, $C_5N$ and $C_4H^-$, which are over-produced in our models at all times. For $l\text{-}C_3H$ and $C_8H^-$, agreement with the observations is obtained only for a C/O ratio = 0.7 with the model using the new network. For $l\text{-}C_3H_2$, only the simulated abundance at early times (before $7\times10^4$ yr) agrees with the observations for the old or new networks and for both C/O ratios. The overall agreement for the long carbon chain containing molecules is much better with the new network for both C/O ratios, the model producing lower abundances of long chain.
The situation is very different in L134N (N). In this dark cloud, carbon chains are much less abundant than in TMC-1 (CP) and many species are overproduced in our model at all ages. The species not reproduced at all are: l-$C_3H$, c-$C_3H$, CN, $C_3N$, $HC_3N$, $C_5H$ and l-$C_3H_2$. For the species $C_6H^-$, $HC_5N$, $HC_7N$ and $C_4H$, the new rate constants predict abundances closer to the observations. It is clear that the observed abundances of both TMC-1 (CP) and L134N(N) cannot be simulated with a unique set of physical parameters and chemical conditions.

### 3.3.2 General agreement between the model and the observations

To quantify the general agreement between the model predictions and the observations, we can use several methods. For a pure gas-phase model, uncertainties in the calculated model abundances can be relatively easily obtained by studying the propagation of model parameter uncertainties (Wakelam *et al.* 2010). For a chemical model combining gas phase and grain surface chemistries, such estimations are more complicated and such studies have not yet been undertaken. Instead, we will use methods similar to those described in Wakelam *et al.* 2006 and Garrod *et al.* 2007. In Wakelam et al. 2006, a "distance of disagreement" is computed for each species and at each time step. This distance corresponds to the difference between the observed and modeled abundances: $\left|\log(X_i) - \log(X_{obs,i})\right|$, with $X_i$ being the modeled abundance of species i and $X_{obs,i}$ the observed abundance. Fig. 2 shows the mean "distance of disagreement" for the two clouds as a function of time. In this case, a smaller value for the distance of disagreement corresponds to a better agreement. Using this method, all observed species have the same weight on the agreement and only the observed species listed in Table 3 were compared to the model results. Upper limits have been removed. In total, 23 species are compared for TMC-1 and 12 in L134N. In the case of TMC-1, several published data exist for some molecules with large differences. For the figures presented in this paper, we have chosen in each case the observed abundance that matches best the model. For TMC-1, both comparison methods give the same result: the new network improves the agreement and both C/O ratios produce a reasonable agreement at similar times. There are in fact two values for the age which show better agreement. The first one equal to $10^5$ yr corresponds to a time which is dominated by gas-phase chemistry and the second one, around $1\text{-}2\times10^6$ yr, is heavily influenced by grain surface chemistry and gas-grain interactions. Using the observed abundances that do not match best the models strongly reduces the general agreement but does not change the fact that the new network improves the agreement. For L134N, there are also two age determinations with the new network corresponding to gas phase and grain surface productions. The first minimum, around $3\text{-}5\text{x}10^4$ yr for a C/O ratio of 0.7, corresponds to a less evolved cloud with containing large free C, N and O abundances in the gas-phase and correspondingly low carbon chain concentrations. This age agrees relatively well with the CO abundance, but is not compatible with $NH_3$, $N_2H^+$, NO and OH abundances. The second age determination which is likely to be the more reliable one, around $6\text{x}10^5$ yr, corresponds to a time where strong depletion effects are predicted to occur and grain chemistry begins to play an important role. The new network does not have a strong effect although the older network seems to produce a slightly better agreement with observations. In either case, for L134N the model using the C/O ratio of 0.7 gives clearly better results than the larger C/O ratio.

For both clouds, if we consider more molecules in the comparison such as sulfur bearing species and oxygen rich molecules (see Tables 3 and 4 in Garrod *et al.* 2007), the general agreement is smaller but the conclusions on the best model and ages are the same.

## 4 Conclusions

In this paper, we have reviewed a large number of reactions involved in the formation and destruction of carbon chains under dense cloud conditions. Incoherent rate constants and branching ratios used in the model have been updated using experimental and theoretical literature data. In total 8 new species are introduced, 41 new reactions are proposed and 122 rate constants from kida.uva.2011 are changed. Looking at the effect of these changes in the model predictions for dense clouds, we find that the new rates, particularly through the increase of the rate constants for O atom reactions, have a strong effect on carbon chains, significantly decreasing the simulated long chain abundances. The comparison of the model predictions and observed abundances in two well-studied dark clouds TMC-1 (CP) and L134N (N) shows that the new updated chemical network reproduces better the observations in TMC-1 (CP) whichever C/O elemental ratio that is used. For L134N (N), the observed abundances are equally reproduced by the older and new chemical network but only for C/O = 0.7. Finally, this study shows the crucial need for the experimental determination of rate constants for the O + $C_nH$ and N + $C_nH$ reactions, particularly at low temperature.

All the reactions and rate constants discussed in this paper will be included in the online KInetic Database for Astrochemistry (KIDA, http://kida.obs.u-bordeaux1.fr/) with their associated datasheets and we encourage astrophysicists to include these updated values in their models.

Acknowledgements:
The authors thank the following funding agencies for their partial support of this work: the French CNRS/INSU program PCMI and the Observatoire Aquitain des Sciences de l’Univers.

## Annex: chemical review

## Chemistry overview:

The **$^1C_{2n+1}$ species** ($C_3$, $C_7$, …) are characterized by a singlet ground state. They have a "closed shell like" structure. The reaction of atomic carbon with $C_3$ does not show any barrier (Wakelam et al. 2009) but there is no data for the reactions of atomic oxygen and nitrogen. Using semi-empirical prediction criteria (Smith *et al.* 2006) leads to a barrier for the O + $C_3$ reaction which is also very likely to be the case for the N + $C_3$ reaction as atomic nitrogen is somewhat less reactive than atomic oxygen with hydrocarbons. We also take in account the low reactivity of $^1C_3$ with hydrocarbons and NO (Nelson *et al.* 1982, Nelson *et al.* 1981, Gu *et al.* 2007, Guo *et al.* 2007, Li *et al.* 2005, Mebel *et al.* 2007a). Consequently, we consider that $^1C_{2n+1}$ molecules are non-reactive with N atoms, non-reactive with O atoms (albeit with a large uncertainty) but reactive with carbon atoms.

The **$^3C_{2n}$ radicals** are characterized by triplet ground states with low lying singlet excited states (+ 0.1 eV) (except $C_2$ for which the singlet state is the ground state) (Martin & Taylor 1995). Their reactivity is deduced from the low energy excited state of $^3C_2$ which is reactive with closed shell molecules (Reisler *et al.* 1980a, Reisler *et al.* 1980b, Paramo *et al.* 2008, Daugey *et al.* 2008) and with O and N atoms (Becker *et al.* 2000). Moreover, considering the high reactivity of atomic carbon there is very likely to be no barrier in the entrance valley for all the C + $^3C_{2n}$ reactions. As a result, $^3C_{2n}$ radicals are considered here to be reactive (no barrier in the entrance valley) with C, N and O atoms. We present the reactivity of $C_2$ separately as the ground state is different from the other $C_{2n}$ species. The $^1C_2$ radical has a very high reactivity even with saturated molecules such as alkanes (Paramo et al. 2008). As a result, there is little doubt that the reactions of C, N and O atoms with $^1C_2$ have no barriers in the entrance valley.

The **$^2C_nH$ radicals** are characterized by doublet ground states and are very reactive species. The CH radical is reactive with molecules (Canosa *et al.* 1997, Daugey *et al.* 2005, Loison *et al.* 2006, Loison & Bergeat 2009, Goulay *et al.* 2009, Blitz *et al.* 1999, Blitz *et al.* 1997, Blitz *et al.* 2012) and atoms (Brownsword *et al.* 1996, Messing *et al.* 1980). The $C_2H$ radical is also reactive with molecules (Pedersen *et al.* 1993, Goulay *et al.* 2011, Woon 2006) and atoms (Boullart *et al.* 1996, Devriendt *et al.* 1996). The $C_4H$ radical is known to be reactive with molecules (Berteloite *et al.* 2010a, Berteloite *et al.* 2010b). There is very little doubt therefore, that $^2C_nH$ radicals are reactive (for both odd and even n) with atomic radicals C, N, and O. Even if there are cyclic and linear isomers for $C_nH$ when n>2, we consider for this study only cyclic isomers for the $C_3H$ radical. As both linear and cyclic forms are radicals in ground doublet states, they are therefore likely to possess similar reactivities.

The **$C_nH_2$ radical** family is complex. For odd and even n there is the usual linear form $^1HC_nH$ (the even n form is by far the most stable), the cumulenic form $^1H_2C_n$ with the two H atoms on the terminal carbon atom H$_2$C=C=C..C| (although this is not a particularly stable form for either odd or even n it is nonetheless the one detected in dense molecular clouds for n = 3,4,6) (Cernicharo *et al.* 1991, Langer *et al.* 1997), a linear triplet form for odd n, $^3HC_{2n+1}H$ and various stable cyclic forms in a singlet state for odd n, for example 10 isomers have been theoretically identified for $C_5H_2$ (Mebel et al. 2007a). In this study we consider only the linear $^1C_nH_2$ isomers for n>3 and the $C_3H_2$ case will be studied in detail in a forthcoming paper. The reactivity of $^1C_nH_2$ is detailed in the Annex. Not all the $^1C_nH_2$ isomers are likely to be reactive with N and C atoms. The $^1C_nH_2$ are considered to be reactive with O atoms except for $^1C_2H_2$ and $^1C_4H_2$.

The $^1$**$C_{2n+1}O$ molecules** are very stable molecule and should be present in molecular clouds as they are likely to be a (minor) product of the $C_{2n+1}H$ + O reactions. (Zhao *et al.* 2007) There are no experimental or theoretical studies of their reactivity to our knowledge, but they have "closed shell like" structure and should only react with carbon atoms.

The $^2$**$C_nN$ radicals** are characterized by doublet ground states and are very reactive species. The CN radical is reactive with molecules (Sims *et al.* 1993, Gannon *et al.* 2007, Georgievskii & Klippenstein 2007, Meads *et al.* 1993) and atoms (Daranlot et al. 2012, Whyte & Phillips 1983, Schacke *et al.* 1973, Albers *et al.* 1975, Schmatjko & Wolfrum 1978, Schmatjko & Wolfrum 1977, Titarchuk & Halpern 1995). The $C_2N$ radical is also reactive with molecules (Zhu *et al.* 2003, Wang *et al.* 2005, Wang *et al.* 2006). There is therefore little doubt that all $C_nN$ radicals react rapidly with C, N, and O atoms. On the product side, there are currently some inconsistencies in current astrochemical databases. Here, we find that some tricks have been used to avoid the introduction of new species. For example, the products of the O + $C_6N$ reaction are currently given as OCN + $C_6$ ($\Delta_rH_{298}$ = -30 kJ/mol) instead of the preferred ones CO + $C_6N$ ($\Delta_rH_{298}$ = -474 kJ/mol). Similarly, the reaction N + $C_6H$ leads to CN + $C_5H$ production ($\Delta_rH_{298}$ = -118 kJ/mol) with a low rate constant instead of H + $C_6N$ ($\Delta_rH_{298}$ = -227 kJ/mol) production with a large rate constant. The reaction products for both of these reactions have been modified from the ones we would expect to avoid the introduction of a new species, $C_6N$. In contrast, we consider that $C_6N$, $C_8N$ and $C_{10}N$ species are required to get a better description of carbon chains in the models and these species have therefore been introduced and the chemistry completed, including the main ionic and DR reactions.

The $^1$**$HC_{2n+1}N$ family** are very stable molecules. They can be formed by H + $C_{2n+1}N^-$, CN + $C_{2n}H_2$ reactions and DR of $HC_{2n+1}NH^+$. $HC_3N$ reacts with carbon atoms (Li *et al.* 2006) but not with N and O atoms. We extrapolate these results to the $HC_{2n+1}N$ family. The rate constant for the C + $HC_{2n+1}N$ reaction is deduced from the C + alkynes rate constants (Chastaing *et al.* 1999) taking into account the back dissociation (Li et al. 2006). We do not introduce the $HC_{2n}N$ neutral family nor the $H_2C_nN$ family (but $H_2CN$ and $H_2C_2N$ are already present in KIDA) as these molecules are not efficiency synthesized in dense molecular clouds.

### $^3C$ + $^1C_{2n+1}$ reactions:

The reactants correlate with 3 triplet states, $^3A'$ + 2 $^3A''$, and the products, given in Table 1, correlate only with a singlet electronic state. By comparison with the reactions of C + alkenes and alkynes (Haider & Husain 1992, Haider & Husain 1993a, Haider & Husain 1993b, Husain & Ioannou 1997, Husain & Kirsch 1971b, Husain & Kirsch 1971a, Clary et al. 1994, Takahashi & Yamashita 1996, Chastaing et al. 1999, Chastaing et al. 2001, Bergeat & Loison 2001, Chastaing et al. 2000, Loison & Bergeat 2004), there is very likely to be no barrier for one triplet surface. Product formation requires intersystem crossing as the only exothermic products are those formed in singlet states. By comparison with the C + $C_2H_2$ reaction (Bergeat & Loison 2001, Mebel *et al.* 2007b, Costes *et al.* 2009), there is no doubt that triplet-singlet inter system crossing is efficient for these reactions, competing with back dissociation. We estimate the rate constant for these reactions by comparison with the C($^3P$) + $^1C_2H_2$, $^1C_2H_4$, $^1CH_3CCH$ (methylacetylene) and $^1CH_2CCH_2$ (allene) reactions (Chastaing et al. 1999, Chastaing et al. 2001), considering that for the C + $C_5$ reaction there is 50 % of triplet adduct back dissociation and 50 % of triplet-singlet crossing only. For the C + $C_7$ reaction we consider that back dissociation represents only 20% and is negligible for the C + $C_9$ reaction.

**$^3$C + $^1C_2$ reactions:**

This reaction has been already studied (Wakelam et al. 2009). The only product is $C_3$ through radiative association.

**$^3$C + $^3C_{2n}$ reactions:**

$^3$C + $^3C_{2n}$ correlate adiabatically with singlet, triplet and quintuplet states. Correlation is complex here, there are 27 surfaces ($^{1,3,5}$A' and 2×$^{1,3,5}$A") and if we consider Ω as a good quantum number, the J = 0 level of carbon atom should correlate with three surfaces, only one (the singlet surface) being reactive. However there are two other singlet surfaces (arising from J=1 or J=2), which may also be attractive (the three surfaces corresponding to $^1$A' and 2 $^1$A"). We recommend a rate constant only at 10 K, with a value close to the ones obtained for the C + alkenes and alkynes reactions multiplied by the population of J = 0 at 10 K (equal to 0.78): k(10K) = $2.4\times10^{-10}$ cm$^3$ molecule$^{-1}$ s$^{-1}$ with $F_0$ = 3 as there are no measurements of these reactions. It should be noted that the temperature dependence for the rate constants of these reactions may be similar to the $^3$C + $^3O_2$ reaction studied in CRESU experiments: k(T) = $4.7\times10^{-11}\times(T/300)^{-0.34}$ cm$^3$ molecule$^{-1}$ s$^{-1}$. (Geppert *et al.* 2000)

**$^3$C + $^2C_nH$ reactions:**

$^2C_nH$ radicals are either in a $^2\Pi$ ground state (CH) or in a $^2\Sigma$ ($^2$A') ($C_2H$, $C_3H$, $C_4H$, …) ground state. $^3$C + $C_nH(^2\Pi)$ reagents correlate adiabatically with six doublet states and six quadruplet states and $^3$C + $C_nH$($^2\Sigma$ or $^2$A') correlate adiabatically with three doublet states and three quadruplet states. The product state correlations are complicated due to the exothermicity of these reactions and the presence of low lying excited state of the products. However, as for C + $C_n$ reactions, we always consider the surface arising from the J = 0 level of the carbon atom as barrierless. A reasonable approximation of the rate constant is given by the measured rate constant values of C + alkenes and alkynes reactions multiplied by the population of J = 0 at 10 K (equal to 0.78): k(10K) = $2.4\times10^{-10}$ cm$^3$ molecule$^{-1}$ s$^{-1}$ with $F_0$ = 3.0 except for C + CH ($F_0$ = 2.0) for which a theoretical study has been performed (Boggio-Pasqua *et al.* 2000).

**$^3$C + $^1C_nH_2$ reactions:**

These reactions are very similar to the carbon atoms reactions with alkenes, alkynes, dienes and diynes (Haider & Husain 1992, Haider & Husain 1993a, Haider & Husain 1993b, Husain & Ioannou 1997, Husain & Kirsch 1971b, Husain & Kirsch 1971a, Clary et al. 1994, Chastaing *et al.* 1998, Chastaing et al. 2000). There is very likely to be no barrier in the entrance valley for these reactions. The rate constant for these reactions should be very close to the ones for C + alkenes, alkynes, dienes and diynes. We recommend a value similar to the ones for the C + $C_2H_2$ and C + $C_3H_4$ reactions (Chastaing et al. 1998, Chastaing et al. 2000) .

**$^3$C + $^1C_{2n+1}O$ reactions:**

There is likely to be no barrier in the entrance valley for these reactions by comparison with the C + alkenes and alkynes reactions. This case is very similar to the C + $C_{2n+1}$ one (see section 3.1.1). We reach the same conclusions and recommend k(10K) = $3.0\times10^{-10}$ cm$^3$ molecule$^{-1}$ s$^{-1}$ with $F_0$ = 1.6.

**$^3$C + $^2C_nN$ reactions:**

$^2C_nN$ radicals are isoelectronic with $^2C_nH$ radicals (either in a $^2\Pi$ ground state or in a $^2\Sigma$ ($^2$A') ground state). As a result, we consider that $^2C_nN$ radicals have similar reactivity to $^2C_nH$

radicals leading to a recommended rate constant k(10K) = $2.4\times10^{-10}$ $cm^3$ $molecule^{-1}$ $s^{-1}$ (with $F_0$ = 3.0 as there are neither measurements nor theoretical studies for these reactions).

**$^3C$ + $^1HC_{2n+1}N$ reactions:**
These reactions are very similar to the $^3C$ + $^1C_{2n+1}$, $^1C_nH_2$, $^1C_{2n+1}O$ ones. There is likely to be no barrier in the entrance valley as shown by (Li et al. 2006) for C + $HC_3N$. As the exit channel on the triplet surface is only -30 to -50 kJ/mol below the reactant C + $HC_{2n+1}N$ level, back dissociation is calculated to be equal to 60% for C + $HC_3N$ (Li et al. 2006) leading to a rate constant smaller than the capture one, close to k(10K) = $1.0\times10^{-10}$ $cm^3$ $molecule^{-1}$ $s^{-1}$.

**$^4N$ + $^3C_{2n}$ reactions:**
Considering the value of the rate constant for the N + $^3C_2$ reaction, measured to be $2.8\times10^{-11}$ $cm^3$ $molecule^{-1}$ $s^{-1}$ (Becker et al. 2000), we consider no barrier in the entrance valley for these reactions. $^4N$ + $C_2(a^3\Pi)$ reagents correlate adiabatically with two doublet, two quadruplet and two sextuplet states. Considering that only the doublet surfaces have no barrier, there is an electronic degeneracy factor equal to 4/24 = 1/6 leading to a capture rate constant close to $9\times10^{-11}\times(T/300)^{0.17}$ $cm^3$ $molecule^{-1}$ $s^{-1}$. The capture rate constant seems overestimated by comparison with the measured value equal to $2.8\times10^{-11}$ $cm^3$ $molecule^{-1}$ $s^{-1}$ (Becker et al. 2000). As a result, we recommend a rate constant similar to the N + $^3C_2$ ones for the N + $^3C_{2n}$ reactions: k(T) = $3\times10^{-11}\times(T/300)^{0.17}$ $cm^3$ $molecule^{-1}$ $s^{-1}$ with $F_0$ = 3.0.

**$^4N$ + $^1C_2$ reaction:**
The reactants $^4N$ + $^1C_2$ correlate adiabatically with quadruplet states, as do the products: $^2CN$ + $^3C$. As a result, there is no electronic degeneracy factor leading to a high capture rate constant close to $5\times10^{-10}\times(T/300)^{0.17}$ $cm^3$ $molecule^{-1}$ $s^{-1}$. However by comparison with the N + $^3C_2$ reaction (Becker et al. 2000) where the rate constant is notably smaller than the capture rate value we recommend a lower value for the rate constant: k(T) = $2\times10^{-10}\times(T/300)^{0.17}$ $cm^3$ $molecule^{-1}$ $s^{-1}$ with $F_0$ = 3.0.

**$^4N$ + $^2C_nH$ reactions:**
$^4N$ + $C_nH(^2\Sigma$ or $^2A')$ reagents correlate adiabatically with one triplet state and one quintuplet state and $^4N$ + $C_nH(^2\Pi)$ correlate adiabatically with two triplet and two quintuplet states. Considering that only the triplet surfaces have no barrier and that both triplet surfaces are reactive as excited $CN(A^2\Pi)$ states are accessible in the case of $C_nH(^2\Pi)$, there is an electronic degeneracy factor equal to 3/8. The capture rate constant including electronic degeneracy for N + CH is equal to $(3/8)\times4\times10^{-10}\times(T/300)^{0.17}$ $cm^3$ $molecule^{-1}$ $s^{-1}$ = $1.5\times10^{-10}\times(T/300)^{0.17}$ $cm^3$ $molecule^{-1}$ $s^{-1}$, in good agreement with the experimental value for the N + CH reaction between 216 and 584K: $1.6\times10^{-10}\times(T/300)^{-0.09}$ $cm^3$ $molecule^{-1}$ $s^{-1}$ (Brownsword et al. 1996). For the $^4N$ + $C_nH$ reactions we recommend the experimental room temperature rate constant (Brownsword et al. 1996) with the capture rate theory predicting a positive temperature dependence: $1.6\times10^{-10}\times(T/300)^{0.17}$ $cm^3$ $molecule^{-1}$ $s^{-1}$. It should be noted that a new experimental and theoretical study between 10 K and 296 K has just been published showing a positive temperature dependence and a smaller rate constant at low temperature (Daranlot *et al.* 2013). The value for this rate constant will be updated in a future study.

**$^4N$ + $^1C_nH_2$ reactions:**
There are various measurements at room temperature for the N + $C_2H_2$ reaction leading to a very low rate constant (Sato *et al.* 1979, Herron & Huie 1968) and an activation barrier of 86 kJ/mol has been theoretically predicted by (Balucani *et al.* 2000). There is therefore likely to be a notable barrier for all these reactions involving $^1C_{2n}H_2$ leading to negligible values of the rate constant for these reactions at low temperature. In this study the reactivity of $^1C_{2n+1}H_2$ with N atoms is neglected.

**$^4$N + $^2$C$_n$N reactions:**

$^4N + C_nN(^2\Sigma$ or $^2A')$ reagents correlate adiabatically with one triplet and one quintuplet states and $^4N + C_nN(^2\Pi)$ correlate adiabatically with two triplet surfaces and two quintuplet states. Considering that only the triplet surfaces have no barrier and that both triplet surfaces are reactive as excited $CN(A^2\Pi)$ states are accessible in the case of $C_nN(^2\Pi)$, there is an electronic degeneracy factor equal to 3/8. The capture rate constant including electronic degeneracy for N + CN is equal to $(3/8)\times4\times10^{-10}\times(T/300)^{0.17}$ cm$^3$ molecule$^{-1}$ s$^{-1}$ = $1.5\times10^{-10}\times(T/300)^{0.17}$ cm$^3$ molecule$^{-1}$ s$^{-1}$, slightly above the N + CN experimental rate constant: $9\times10^{-11}\times(T/300)^{0.42}$ cm$^3$ molecule$^{-1}$ s$^{-1}$ between 56 and 300K. There is very likely to be no barrier for these reactions and as there is very little data on these reactions for n>1 (Whyte & Phillips 1983), we recommend the usual capture rate temperature dependence for the N + $C_nN$ reactions (for n>1) with a value at 300 K equal to the experimentally measured one for the N + CN reaction. For the N + CN reaction we recommend the measured values of Daranlot *et al.* 2012a:

$k(T) = 9\times10^{-11}\times(T/300)^{0.42}$ cm$^3$ molecule$^{-1}$ s$^{-1}$ for N + CN reaction

$k(T) = 9\times10^{-11}\times(T/300)^{0.17}$ cm$^3$ molecule$^{-1}$ s$^{-1}$ for N + $C_{n>1}N$ reaction

**$^3$O + $^1$C$_2$ reaction:**

$^3O + {^1C_2}(X^1\Sigma^+)$ reagents correlate adiabatically with triplet states as do the products $^1CO + {^3C}$, so the electronic degeneracy factor is 1. This case is similar to the O + butenes one where reaction is thought to proceed through addition via Van der Waals complex formation. The low temperature reactivity of the O + alkenes reactions have been studied in CRESU experiments (Sabbah *et al.* 2007) showing that for propene, 1-butene, iso and trans butenes, the low rate constant value at room temperature was due to a submerged barrier and that the rate constant at low temperature was close, at least for iso and trans butenes, to the capture limited rate value. However, as the capture rate constant is dominated by the dispersion term which is proportional to $\alpha^{1/3}$ where $\alpha$ is the polarisability, the lower $\alpha$ for oxygen compared with carbon leads to a smaller rate constant at low temperature for the O + alkenes reactions ($k(25K) = 2\times10^{-10}$ cm$^3$ molecule$^{-1}$ s$^{-1}$ for O + trans-butene) (Sabbah et al. 2007) compared with $k(25K) = 4\times10^{-10}$ cm$^3$ molecule$^{-1}$ s$^{-1}$ for C + allene/methyl-acetylene (Chastaing et al. 2000)). For the $^3O + {^1C_2}$ reaction there is very likely to be no submerged barrier leading to a rate constant close to $2.0\times10^{-10}$ cm$^3$ molecule$^{-1}$ s$^{-1}$ in the 10-300 K range using the same capture rate constant deduced from experimental studies of O + alkenes reactions. (Sabbah *et al.* 2007). We recommend an uncertainty factor equal to 3.0 as the reference reactions, O + alkenes, show a complex temperature dependence due to submerged barrier leading to increased uncertainty of the rate constant value at 10K.

**$^3$O + $^3$C$_{2n}$ reactions:**

Considering the reactivity of oxygen atoms with alkenes (Sabbah *et al.* 2007) as well as the large rate constant for the O + $^3C_2$ reaction ($k(300K) = 1.0\times10^{-10}$ cm$^3$ molecule$^{-1}$ s$^{-1}$ (Becker et al. 2000)), there is very likely to be no barrier for these reactions. $^3O + {^3C_{2n}}$ reactants correlate adiabatically with singlet, triplet and quintuplet states. The reactions with oxygen atoms are very exothermic so that singlet and triplet state products are accessible. Considering only the singlet surface as reactive leads to an electronic degeneracy factor of 1/9, and considering singlet and triplet surfaces as reactive leads to an electronic degeneracy factor of 4/9. The rate constant for the O + $^3C_2$ reaction ($\Delta_rH$ = 582 kJ/mol) measured by (Becker et al. 2000), corresponds to an electronic degeneracy factor of 4/9 if $k_{capture}$ is equal to $2\times10^{-10}$ cm$^3$

molecule$^{-1}$ s$^{-1}$ suggesting that products in their excited electronic states are produced. As all the O + $^3C_{2n}$ reactions are also highly exothermic, products in their triplet electronic states are likely to be accessible. Then the experimental rate constant of the O + $^3C_2$ reaction at room temperature is recommended for all the O + $^3C_{2n}$ reactions with the theoretical capture rate temperature dependence ($T^{0.17}$). As there is a direct reliable measurement of the O + $^3C_2$ reaction, the uncertainty factor $F_0$ is taken to be equal to 2.

**$^3$O + $^2C_nH$ reactions:**

$^3$O + $C_nH(^2\Sigma$ or $^2A')$ reagents correlate adiabatically with three doublet and three quadruplet states and $^3$O + $C_nH(^2\Pi)$ correlate adiabatically with six doublet and six quadruplet states. We consider that all the doublet surfaces have no barrier (as in the case of the O + $C_2H$ reaction (Georgievskii & Klippenstein 2011)) and correlate with the products in their ground or excited states (excited CO states or $C_nO$ products are energetically accessible). For oxygen atoms, the lowest fine structure level is J = 2, which is fivefold degenerate, leading always to a small temperature dependence due to the evolution of the fine structure population. Neglecting this small temperature dependence and considering only the doublet surfaces as reactive leads to an electronic degeneracy factor of 1/3. Assuming a long range potential for the interaction of oxygen atoms with these radicals similar to the one for O with alkenes and using the (Sabbah et al. 2007) results leads to k(10-300K) = $2.0\times10^{-10}\times$ (2/6) = $7\times10^{-11}$ cm$^3$ molecule$^{-1}$ s$^{-1}$. This proposition is in relatively good agreement with the available rate constant measurements for the O + CH reaction, k(300K) = $9.3\times10^{-11}$ cm$^3$ molecule$^{-1}$ s$^{-1}$ (Messing et al. 1980, Messing *et al.* 1981), the O + $C_2H$ reaction, k(300K) = $9.0\times10^{-11}$ cm$^3$ molecule$^{-1}$ s$^{-1}$ (Boullart et al. 1996, Devriendt et al. 1996) and also with theoretical calculations for the O + CH reaction, k(T) = $8.5\times10^{-11}\times(T/300)^{0.15}$ cm$^3$ molecule$^{-1}$ s$^{-1}$ (Phillips 1990), even if these rates are noticeably smaller than recent statistical calculations for the O + $C_2H$ reaction (Georgievskii & Klippenstein 2011). In contrast to the O + alkenes reactions, there is likely to be no inner barrier for the O + $^2C_nH$ reactions as they are radical-radical reactions, leading to little temperature dependence (through a convolution of the $T^{0.17}$ dispersion dependence with the variation of population of the J = 2 level of the oxygen atom). Consequently, we propose a constant value in the 10-300K range. For all the O + $^2C_nH$ reactions we recommend a constant value deduced from low temperature measurements of the O + alkenes reactions corrected by the appropriate electronic degeneracy factor, leading to k(T) = $7\times10^{-11}$ cm$^3$ molecule$^{-1}$ s$^{-1}$. As there are various scattered measurements and calculations we estimate the uncertainty factor $F_0$ to be equal to 2.0. This value is markedly higher than the previous value used in astrochemical models although it may still be lower than the real value. Clearly experimental measurements, particularly at low temperature, are required for these important reactions.

**$^3$O + $^1C_nH_2$ reactions:**

Temperature dependent rate constants have been measured for the O + $C_4H_2$ reaction (Mitchell et al. 1986) indicating the presence of a barrier. To estimate the rate constant for O + $C_{6,8,10}H_2$ reactions we apply the prediction criteria from Smith et al. 2006 leading to no barrier for the O + $C_{6,8,10}H_2$ reactions (Ionization Energy($C_{6,8,10}H_2$) – Electron Affinity(O) < 8.75 eV, IE ($C_6H_2$) = 9.50 eV (Bieri *et al.* 1977), IE ($C_{8,10}H_2$) < 9.50 eV and EA(O) = 1.439 eV (Joiner *et al.* 2011)). The linear cumulenic $^1C_{5,7,9}H_2$ isomers are also likely to follow the prediction criteria (the calculated IE of $C_5H_2$ is found to be equal to 9.30 eV at M06-2X/cc-pVTZ level). As there is likely to be an inner barrier, similar to the O + alkenes reactions (Sabbah *et al.* 2007), we only recommend a rate constant at 10 K equal to the experimental rate constant (Sabbah et al. 2007) for the reaction of O with iso and trans butenes: k(10K) = $2\times10^{-10}$ cm$^3$ molecule$^{-1}$ s$^{-1}$.

**$^3$O + $^2$C$_n$N reactions:**

The $C_nN$ radicals have similar electronic ground states and polarisabilities as $C_nH$ radicals and we recommend the same rate constant for O + $C_nN$ reactions as for O + $C_nH$ reactions: k(T) = $7\times10^{-11}$ cm$^3$ molecule$^{-1}$ s$^{-1}$ in the 10-300 K range.

**$^1$C$_{2n+1}$O ionic chemistry:**

As we introduce new $C_{2n+1}O$ species we have to describe their ionic chemistry. The ionic chemistry for these species is limited, as $HC_{2n+1}O^+$ ions, formed by the $C_{2n+1}O + H_3^+$, $C_{2n+1}O + HCO^+$, $C_{2n+1}O + HCNH^+$ reactions, do not react with $H_2$ (there are no exothermic pathways) but only with electrons (probably leading to H + $C_{2n+1}O$, $HC_{2n+1}$ + O and $HC_{2n}$ + CO). $HC_3O^+$ is currently present in KIDA and we introduce $HC_{5,7,9}O^+$ cations considering only their formation from $C_{2n+1}O + H_3^+$, $C_{2n+1}O + HCO^+$, $C_{2n+1}O + HCNH^+$ reactions and their loss through Dissociative Recombination (DR) with electrons. The rate constants for the $C_{2n+1}O + H_3^+$, $C_{2n+1}O + HCO^+$, $C_{2n+1}O + HCNH^+$ reactions are calculated following (Woon & Herbst 2009) using calculated polarisabilities and dipole moments from the osu database or they are calculated at the M06-2X/cc-pVTZ level. All data are presented in Table 2. DR rate constants are estimated using the Langevin formula and branching ratios are deduced from similar DR reactions (Florescu-Mitchell & Mitchell 2006).

**$C_{2,4,6,8,10}N^+$ chemistry:**

The $C_{4,6,8,10}N^+$ ions are mainly produced by the $C^+ + HC_{,3,5,7,9}N$ reactions, probably leading to $C_{3,5,7,9}NC^+$ products and not to $C_{4,6,8,10}N^+$ by comparison with $C^+$ + HCN (Harland & McIntosh 1985). However that is not critical as both $C_{4,6,8,10}N^+$ and $C_{3,5,7,9}NC^+$ ions will react with $H_2$ leading to HCN + $C_{3,5,7,9}H^+$ ($\Delta_rH_{298}$ = -268 kJ/mol) and $HCNH^+ + C_{3,8,7,9}$ ($\Delta_rH_{298}$ = -215 kJ/mol). We have to consider $C_2N^+$ as a special case, as the $C^+$ + HCN reaction is supposed to produce $CNC^+$, and $CNC^+$ does not react with $H_2$. (Knight *et al.* 1988) As $C_{4,6,8,10}N^+$ and $C_{3,5,7,9}NC^+$ are very likely to react quickly with $H_2$, the electronic dissociative recombination reactions are not critical. We calculate the capture rate constants following (Woon & Herbst 2009).

**$HC_{2,4,6,8,10}N^+$ chemistry:**

$HC_{2,4,6,8,10}N^+$ ions are mainly formed by quick $H^+$ exchange between the main interstellar ions ($H_3^+$, $HCO^+$, $HCNH^+$) and $C_{2,4,6,8,10}N$ radicals (except for the $C_2N + HCNH^+$ reaction which is endothermic) considering the Proton Affinity values (PA($C_2N$) = 692 kJ/mol, PA($C_4N$) = 789 kJ/mol at M06-2X/cc-p-VTZ and PA($H_2$) = 422 kJ/mol, PA(CO) = 594 kJ/mol and PA(HCN/HNC) = 713/772 kJ/mol from (Hunter & Lias 1998). As a result, the main reactions of $HC_{2,4,6,8,10}N^+$ ions in dense clouds will be with $H_2$. Indeed, using the thermochemistry of $HC_2N^+$ (Scott *et al.* 1999) and $H_2C_2N^+$ (Holmes *et al.* 1993), the $HC_{2,4,6,8,10}N^+ + H_2 \rightarrow H_2C_{2,4,6,8,10}N^+ + H$ reactions are exothermic (-88 kJ/mol for $HC_2N^+ + H_2 \rightarrow H_2C_2N^+ + H$) and considering no barrier for the $HC_{42,4,6,8,10}N^+ + H_2 \rightarrow H_2C_{2,4,6,8,10}N^+ + H$ reactions, we calculate the capture rate constants following (Woon & Herbst 2009).

**$H_2C_{2,4,6,8,10}N^+$ chemistry:**

$H_2C_{2,4,6,8,10}N^+$ ions do not react with $H_2$ as the $H_2C_{2,4,6,8,10}N^+ + H_2 \rightarrow H_3C_{2,4,6,8,10}N^+ + H$ reactions are endothermic (by 268 kJ/mol for $H_3C_2N^+$ (Holmes et al. 1993)). Consequently, the main loss process for $H_2C_{2,4,6,8,10}N^+$ ions is likely to be dissociative recombination, the rate constant having been deduced from similar reactions using (Mitchell et al. 1986) (close to the Langevin values) and (Plessis *et al.* 2012).

Table 1: Summary of reactions review. (Full Table in Supplementary Information)
For carbon atoms reactions T = 10 K only except for C + $C_3$ and C + $C_2H_2$.

| | Reaction | | ΔE kJ/mol | α | β | γ | $F_0$ | g | ref |
|---|---|---|---|---|---|---|---|---|---|
| 1. | $^3C + {}^1C_3$ | $\rightarrow {}^3C_4 + h\nu$ | -495 | 4.0e-14 | -1.0 | 0 | 3 | 0 | (Wakelam *et al.* 2009) |
| 2. | $^3C + {}^3C_4$ | $\rightarrow {}^{1,3}C_2 + {}^1C_3$ | -101 | 2.4e-10 | | | 3 | 0 | (Wakelam et al. 2009) |
| 3. | $^3C + {}^1C_5$ | $\rightarrow {}^1C_3 + {}^1C_3$ | -138 | 1.5e-10 | | | 3 | 0 | We consider an exit Transition State below the entrance channel (Nelson *et al.* 1982) and we assume that k = 0.5×$k_{capture}$ due to back dissociation as it needs triplet-singlet crossing, the $^1C_3 + {}^3C_3$ channel being endothermic by 64 kJ/mol. |
| 4. | $^3C + {}^3C_6$ | $\rightarrow {}^{3,1}C_4 + {}^1C_3$<br>$\rightarrow {}^1C_5 + {}^{1,3}C_2$ | -147<br>-110 | 2.0e-10<br>4.0e-11 | | | 3<br>3 | 0<br>0 | (Diaz-Tendero *et al.* 2006, Wakelam et al. 2009) |
| 5. | $^3C + {}^1C_7$ | $\rightarrow {}^1C_3 + {}^1C_5$ | -131 | 2.0e-10 | | | 3 | 0 | We consider an exit Transition State below the entrance channel (Nelson *et al.* 1982) and we assume that k = 0.8×$k_{capture}$. |
| 6. | $^3C + {}^3C_8$ | $\rightarrow {}^1C_7 + {}^{1,3}C_2$<br>$\rightarrow {}^{3,1}C_6 + {}^1C_3$<br>$\rightarrow {}^1C_5 + {}^3C_4$ | -69<br>-105<br>-80 | 1.0e-11<br>1.6e-10<br>7.0e-11 | | | 3<br>3<br>3 | 0<br>0<br>0 | |
| 7. | $^3C + {}^1C_9$ | $\rightarrow {}^1C_7 + {}^1C_3$<br>$\rightarrow {}^1C_5 + {}^1C_5$ | -116<br>-116 | 2.1e-10<br>0.3e-10 | | | 3<br>3 | 0<br>0 | (Diaz-Tendero et al. 2006, Wakelam et al. 2009). This reaction occurs through triplet-singlet crossing but it involves a large system so the adduct lifetime is long. |
| 8. | $^3C + {}^3C_{10}$ | $\rightarrow {}^{3,1}C_8 + {}^1C_3$<br>$\rightarrow {}^{3,1}C_6 + {}^1C_5$ | -108<br>-90 | 1.8e-10<br>6.0e-11 | | | 3<br>3 | 0<br>0 | The most stable form of $C_{10}$ is the cyclic (c-$^1C_{10}$) one (Van Orden & Saykally 1998) |
| 9. | $^3C + {}^1C_{11}$ | $\rightarrow {}^1C_9 + {}^1C_3$<br>$\rightarrow {}^1C_7 + {}^1C_5$ | | 2.1e-10<br>0.3e-10 | | | 3<br>3 | 0<br>0 | Same as C + $C_9$ |
| | | | | | | | | | |
| 10. | $^3C + CH(X^2\Pi)$ | $\rightarrow {}^{1,3}C_2 + {}^2H$ | -257 | 2.4e-10 | | | 2 | 0 | (Boggio-Pasqua *et al.* 2000) |
| 11. | $^3C + C_2H(X^2\Sigma)$ | $\rightarrow {}^1C_3 + {}^2H$ | -235 | 2.4e-10 | | | 3 | 0 | Same as C + CH |
| 12. | $^3C$ + c-$^2C_3H(X^2A')$ | $\rightarrow {}^3C_4 + {}^2H$<br>$\rightarrow {}^1C_2 + {}^2C_2H$ | -175<br>-41 | 2.4e-10<br>0 | | | 3<br>0 | 0<br>0 | Same as C + CH |

Table 2: Properties of molecules and radicals for rate constant calculations.

| Reaction | $\alpha$ (Å$^3$) | $\mu$(D) | references |
|---|---|---|---|
| $C_3O$ | 6.03 | 2.47 | (Woon & Herbst 2009) |
| $C_5O$ | 10.9 | 3.82 | M06-2X/cc-pVTZ calculations using Gaussian09 |
| HCN | 2.50 | 3.00 | (Woon & Herbst 2009) |
| HNC | 2.73 | 3.09 | (Woon & Herbst 2009) |
| $HC_3N$ | 5.85 | 3.79 | (Woon & Herbst 2009) |
| $HC_5N$ | 9.61 | 4.55 | M06-2X/cc-pVTZ calculations using Gaussian09 |
| $HC_7N$ | 15.46 | 5.12 | M06-2X/cc-pVTZ calculations using Gaussian09 |
| N | 1.100 | 0 | http://cccbdb.nist.gov/ |
| $H_2$ | 0.787 | 0 | http://cccbdb.nist.gov/ |
| $C_2N$ | 4.27 | 0.37 | (Woon & Herbst 2009) |
| $C_4N$ | 7.73 | 0.08 | M06-2X/cc-pVTZ calculations using Gaussian09 |
| $C_6N$ | 14.08 | 0.27 | M06-2X/cc-pVTZ calculations using Gaussian09 |
| $C_8N$ | 22.2 | 0.58 | M06-2X/cc-pVTZ calculations using Gaussian09 |

Table 3: Molecular abundances (relative to $H_2$) observed in TMC-1 and L134N.
Notes: a(b) refers to $a\times10^b$. Unless otherwise indicated abundances correspond to the positions TMC-1 $\alpha_{J2000}$ = $04^h$ $41^m$ $41^s.88$, $\delta_{J2000}$ = +25° $41^m$ $27^s$ (cyanopolyyne peak) and L134N $\alpha_{J2000}$ = $15^h$ $54^m$ $06^s.55$, $\delta_{J2000}$ = -2° $52^m$ $19^s$. Most abundances were derived from observed column densities adopting $N(H_2) = 10^{22}$ $cm^{-2}$ for both TMC-1 and L134N (Goldsmith *et al.* 2007).

| | TMC-1 | | L134N | |
|---|---|---|---|---|
| OH | 1.4(-7)<br>2.6(-7) | (Harju *et al.* 2000)<br>(Suutarinen *et al.* 2011) | 7.5(-8) | (Ohishi *et al.* 1992) |
| $O_2$ | <7.7(-8) | (Pagani et al. 2003) | <1.7(-7) | (Pagani et al. 2003) |
| CO | 1.7(-4) | (Pratap *et al.* 1997) | 8.7(-5) | (Dickens *et al.* 2000) |
| $C_3O$ | 1(-10)<br>1.4(-10) | (Ohishi & Kaifu 1998)<br>(Brown *et al.* 1985) | <5(-11) | (Ohishi *et al.* 1992) |
| NO | 2.7(-8) | (Gerin *et al.* 1993) | 2.0(-7)<br>2.0(-8)<br>8.0(-9) | (Gerin *et al.* 1992)<br>(Akyilmaz *et al.* 2007)<br>(Akyilmaz et al. 2007) |
| CH | 1.6(-8) | (Suutarinen *et al.* 2011) | 1.0(-8) | (Ohishi *et al.* 1992) |
| $C_2H$ | 6.0(-8)<br>2.3(-8)<br>7.2(-9) | (Sakai *et al.* 2010)<br>(Turner *et al.* 1999, Turner *et al.* 2000)<br>(Pratap et al. 1997) | 2.3(-9) | (Dickens et al. 2000) |
| c-$C_3H$ | 1.03(-9)<br>6.0(-9) | (Fossé *et al.* 2001)<br>(Turner et al. 2000) | 4.3(-10) | (Turner et al. 2000) |
| l-$C_3H$ | 8.4(-11)<br>1.05(-9) | (Fossé *et al.* 2001)<br>(Turner et al. 2000) | 1.25(-10) | (Turner et al. 2000) |
| $C_4H$ | 7.1(-8)<br>1.23(-8)<br>1.6(-8)<br>5.7(-8)<br>5.7(-8)<br>6.91(-8) | (Agundez *et al.* 2008)<br>(Turner et al. 2000)<br>(Irvine *et al.* 1981)<br>(Brunken et al. 2007)<br>(Thaddeus et al. 2008)<br>(Cordiner *et al.* 2013) | 1.77(-9) | (Turner et al. 2000) |
| $C_5H$ | 5.07(-10) | (Turner et al. 2000) | <4.96(-11) | (Turner et al. 2000) |
| $C_6H$ | 4.7(-10)<br>8.0(-10)<br>7.5(-10) | (Fossé *et al.* 2001)<br>(Turner et al. 2000)<br>(Brunken *et al.* 2007) | <4.3(-11) | (Gupta *et al.* 2009) |
| $C_8H$ | 4.6(-11) | (Brunken *et al.* 2007) | - | |
| c-$C_3H_2$ | 5.8(-9)<br>4.23(-9) | (Fossé *et al.* 2001)<br>(Turner et al. 2000) | 2.08(-9) | (Turner et al. 2000) |
| l-$C_3H_2$ | 2.1(-10) | (Fossé *et al.* 2001) | 4.2(-11) | (Turner et al. 2000) |
| | 3.36(-10) | (Turner et al. 2000) | | |
| CN | 7.4(-10)<br>2.9(-8) | (Pratap et al. 1997)<br>(Crutcher *et al.* 1984) | 4.8(-10) | (Dickens et al. 2000) |
| $C_3N$ | 6(-10)<br>9.5(-9) | (Ohishi & Kaifu 1998)<br>(Thaddeus *et al.* 2008) | <2(10) | (Ohishi *et al.* 1992) |
| $C_5N$ | 3.1(-11) | (Guélin *et al.* 1998) | - | |
| $HC_3N$ | 1.6(-8)<br>4.4(-9) | (Takano *et al.* 1998)<br>(Pratap et al. 1997) | 4.3(-10) | (Dickens et al. 2000) |
| $HC_5N$ | 4(-9)<br>3.3(-9) | (Ohishi & Kaifu 1998)<br>(Bell *et al.* 1997) | 1(-10) | (Ohishi *et al.* 1992) |
| $HC_7N$ | 1(-9)<br>1.1(-9) | (Ohishi & Kaifu 1998)<br>(Bell *et al.* 1997) | <2(-11) | (Ohishi *et al.* 1992) |
| $HC_9N$ | 5(-10)<br>1.9(-10) | (Ohishi & Kaifu 1998)<br>(Bell *et al.* 1997) | - | |
| $HC_{11}N$ | 2.8(-11) | (Bell *et al.* 1997) | - | |
| $C_4H^-$ | <3.7(-12)<br><2.3(-12)<br>8.0(-13) | (Agundez *et al.* 2008)<br>(Thaddeus *et al.* 2008)<br>(Cordiner et al. 2013) | <3.8(-12) | (Agundez *et al.* 2008) |
| $C_6H^-$ | 1.0(-11) | (Brunken *et al.* 2007) | <1.2(-11) | (Gupta et al. 2009) |
| $C_8H^-$ | 2.1(-12) | (Brunken *et al.* 2007) | - | |
| $C_3N^-$ | <7(-11) | (Thaddeus *et al.* 2008) | - | |

Fig. 1: Abundances of $C_nH$ (n=1-8), $c\text{-}C_3H_2$, $l\text{-}C_3H_2$, CN, $C_3N$, $C_5N$, $HC_nH$ (n=3, 5, 7 and 9), OH, NO, $O_2$, $C_3O$, $C_3N^-$, $C_4H^-$, $C_6H^-$, $C_8H^-$ species as a function of time predicted by the four models: C/O=0.7 and old network (red dashed lines), C/O=0.7 and new network (red solid lines), C/O=0.95 and old network (blue dashed lines), C/O=0.95 and new network (blue solid lines). Horizontal lines represent the abundances observed in TMC-1 (CP) (orange) and L134N (N) (green) and horizontal rectangles represent arbitrary uncertainties. The vertical arrow means that only an upper limit is known.

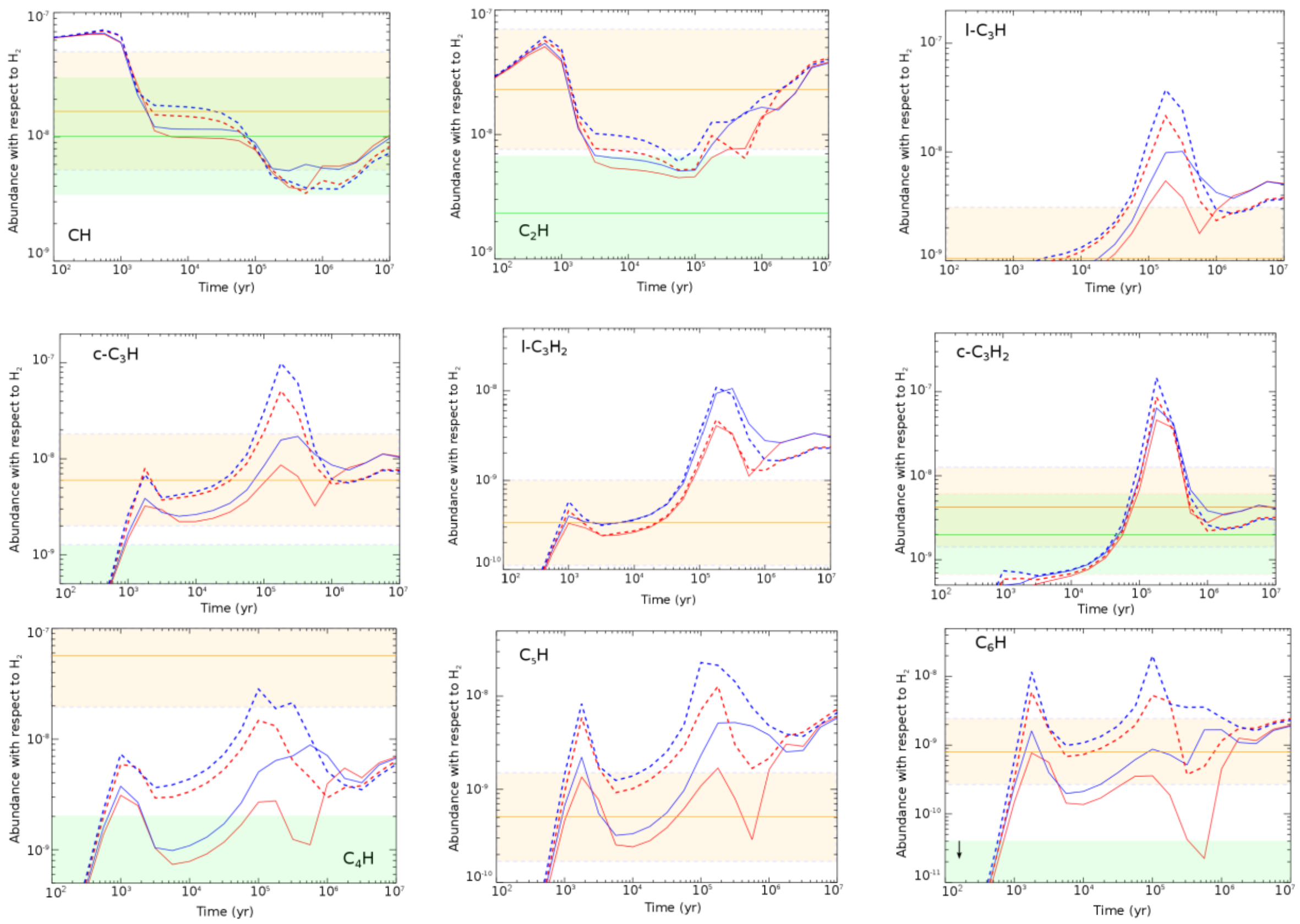

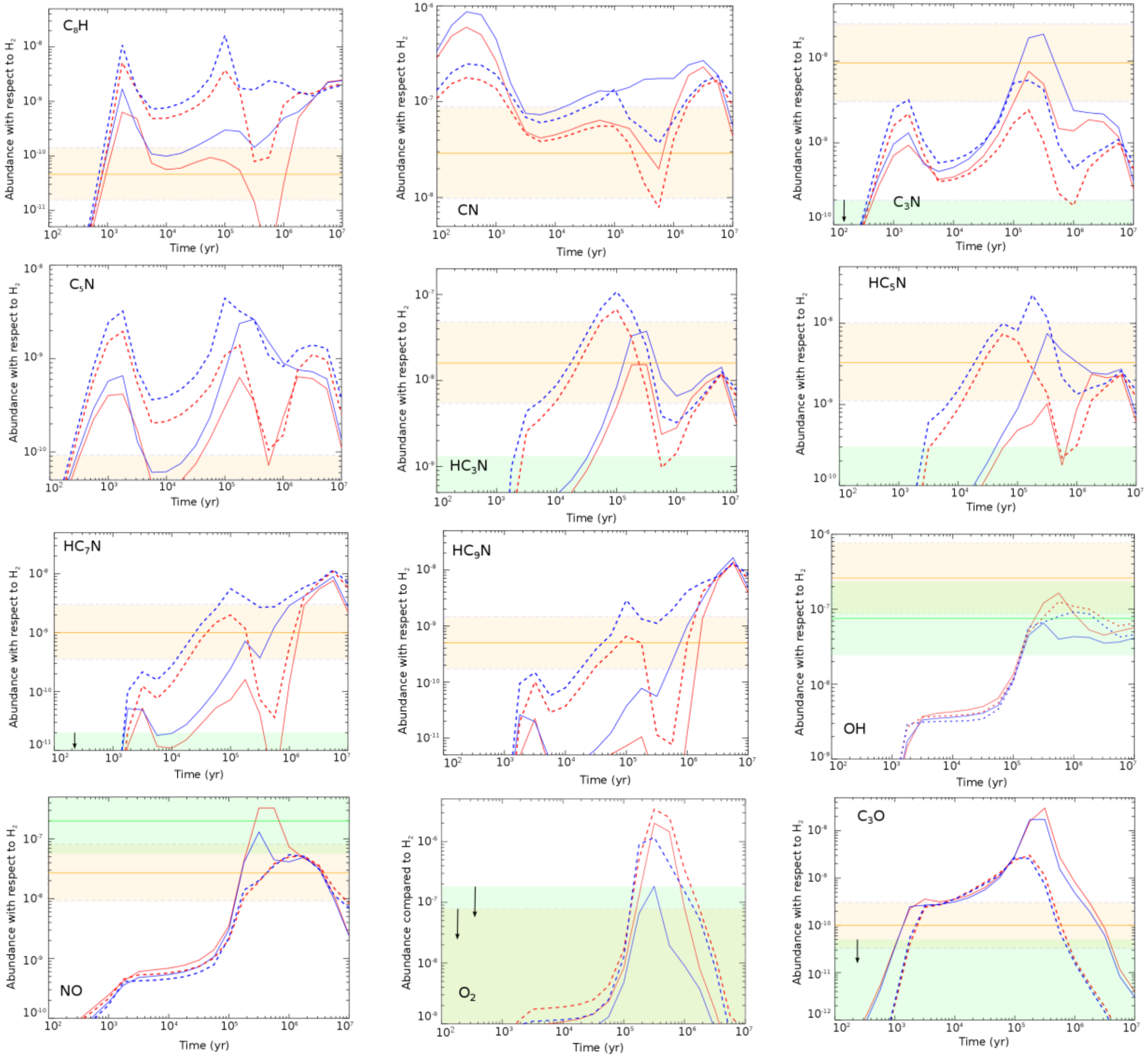

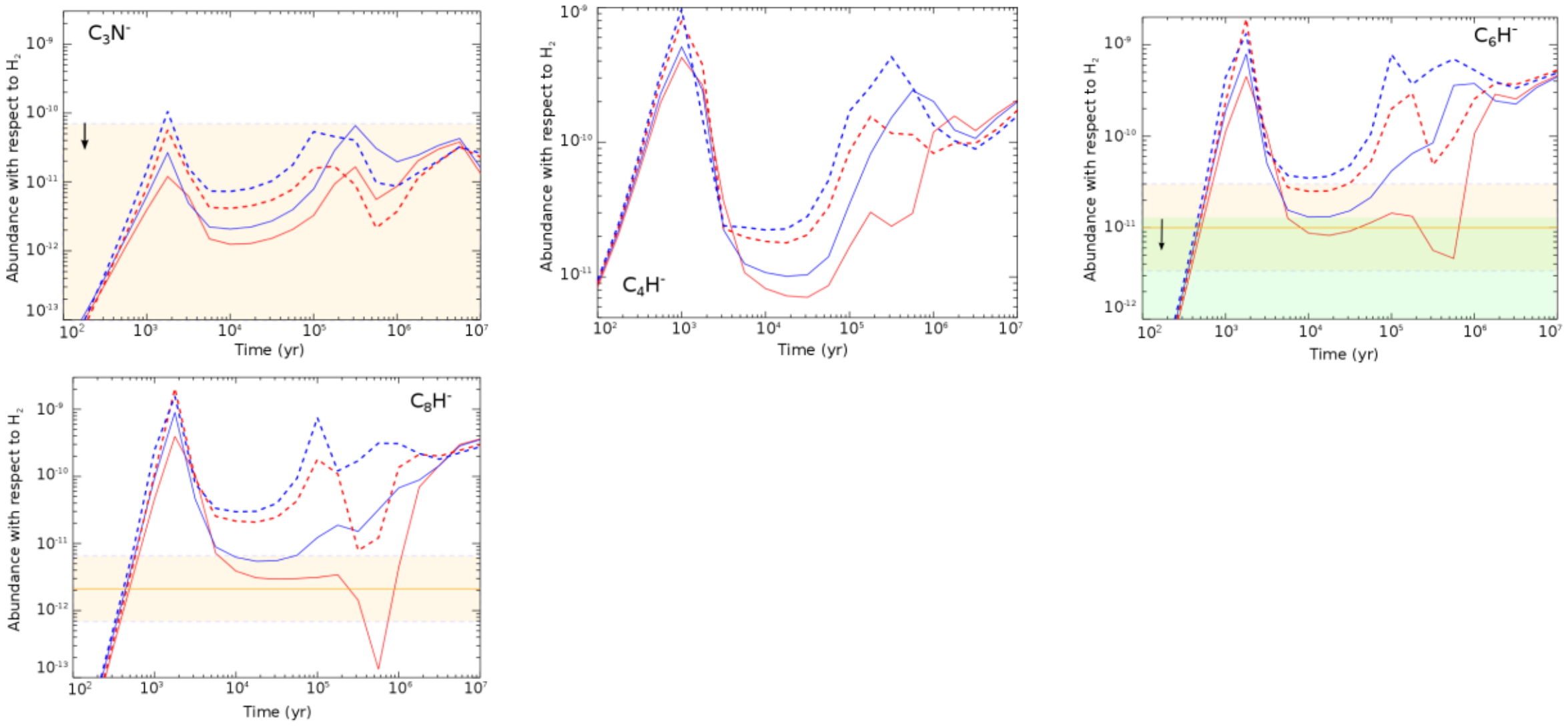

Fig. 2: Mean distance of disagreement computed for the two dark clouds TMC-1 (CP) and L134N (N) predicted by the four models: C/O=0.7 and old network (red dashed lines), C/O=0.7 and new network (red solid lines), C/O=0.95 and old network (blue dashed lines), C/O=0.95 and new network (blue solid lines).

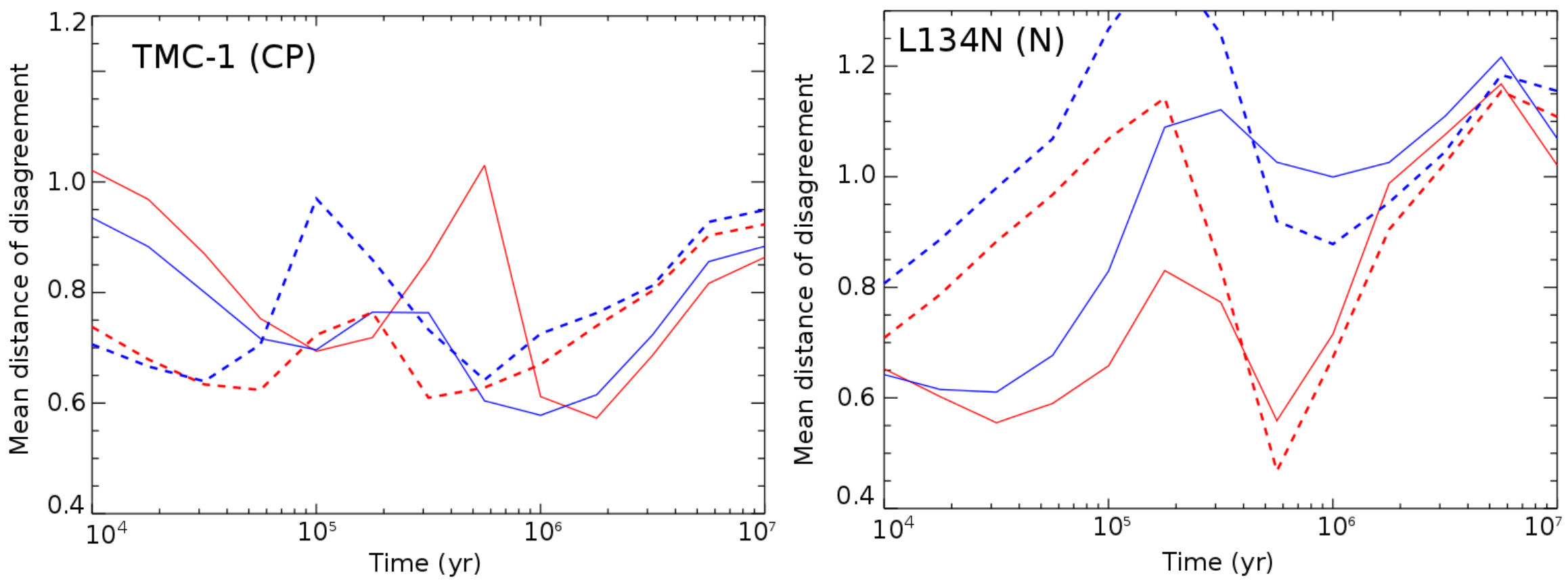

Gaussian 09, M. J. Frisch, G. W. Trucks, H. B. Schlegel, G. E. Scuseria, M. A. Robb, J. R. Cheeseman, J. A. Montgomery, Jr., T. Vreven, K. N. Kudin, J. C. Burant, J. M. Millam, S. S. Iyengar, J. Tomasi, V. Barone, B. Mennucci, M. Cossi, G. Scalmani, N. Rega, G. A. Petersson, H. Nakatsuji, M. Hada, M. Ehara, K. Toyota, R. Fukuda, J. Hasegawa, M. Ishida, T. Nakajima, Y. Honda, O. Kitao, H. Nakai, M. Klene, X. Li, J. E. Knox, H. P. Hratchian, J. B. Cross, V. Bakken, C. Adamo, J. Jaramillo, R. Gomperts, R. E. Stratmann, O. Yazyev, A. J. Austin, R. Cammi, C. Pomelli, J. W. Ochterski, P. Y. Ayala, K. Morokuma, G. A. Voth, P. Salvador, J. J. Dannenberg, V. G. Zakrzewski, S. Dapprich, A. D. Daniels, M. C. Strain, O. Farkas, D. K. Malick, A. D. Rabuck, K. Raghavachari, J. B. Foresman, J. V. Ortiz, Q. Cui, A. G. Baboul, S. Clifford, J. Cioslowski, B. B. Stefanov, G. Liu, A. Liashenko, P. Piskorz, I. Komaromi, R. L. Martin, D. J. Fox, T. Keith, M. A. Al-Laham, C. Y. Peng, A. Nanayakkara, M. Challacombe, P. M. W. Gill, B. Johnson, W. Chen, M. W. Wong, C. Gonzalez, and J. A. Pople, Gaussian, Inc., Wallingford CT, 2009

MOLPRO, version 2009.1, a package of ab initio programs, H.-J.Werner, P. J. Knowles, R. Lindh, F. R. Manby, M. Schütz, P. Celani, T. Korona, A. Mitrushenkov, G. Rauhut, T. B. Adler, R. D.Amos, A. Bernhardsson, A. Berning, D. L. Cooper, M. J. O. Deegan, A. J. Dobbyn, F. Eckert, E. Goll, C. Hampel, G. Hetzer, T. Hrenar, G. Knizia, C. Köppl, Y. Liu, A. W. Lloyd, R. A.Mata, A. J. May, S. J. McNicholas, W. Meyer, M. E. Mura, A. Nicklass, P. Palmieri, K. Pflüger, R. Pitzer, M. Reiher, U. Schumann, H. Stoll, A. J. Stone, R. Tarroni, T. Thorsteinsson, M.Wang, and A. Wolf, see http://www.molpro.net.

Agundez M., Cernicharo J., Guélin M., Gerin M., McCarthy M.C., Thaddeus P., 2008, A&A, 478**,** L19

Akyilmaz M., Flower D.R., Hily-Blant P., Forêts G.P.d., Walmsley C.M., 2007, A&A, 462**,** 221

Albers A.A., Hoyermann K., Schacke H., Schmatko K.J., Wolfrum J., 1975, Proc. 15th Symps. (Intl) Combust.**,** 765

Balucani N., Alagia M., Cartechini L., Casavecchia P., Volpi G.G., Sato K., Takayanagi T., Kurosaki Y., 2000, J. Am. Chem. Soc., 122**,** 4443

Baulch D.L., Bowman C.T., Cobos C.J., Cox R.A., Just T., Kerr J.A., Pilling M.J., Stocker D., Troe J., Tsang W., Walker R.W., Warnatz J., 2005, J. Phys. Chem. Ref. Data, 34**,** 757

Becker K.H., Donner B., Dinis C.M.F., Geiger H., Schmidt F., Wiesen P., 2000, Z. Phys. Chem., 214**,** 503

Bell M.B., Feldman P.A., Travers M.J., McCarthy M.C., Gottlieb C.A., Thaddeus P., 1997, ApJ, 483**,** L61

Bergeat A., Hickson K.M., Daugey N., Caubet P., Costes M., 2009, Phys. Chem. Chem. Phys., 11**,** 8149

Bergeat A., Loison J.-C., 2001, PCCP, 3**,** 2038

Berteloite C., Le Picard S.D., Balucani N., Canosa A., Sims I.R., 2010a, Phys. Chem. Chem. Phys., 12**,** 3666

Berteloite C., Le Picard S.D., Balucani N., Canosa A., Sims I.R., 2010b, Phys. Chem. Chem. Phys., 12**,** 3677

Bieri G., Burger F., Heilbronner E., Maier J.P., 1977, Helv. Chim. Acta, 60**,** 2213

Blitz M., Pesa M., Pilling M.J., Seakins P.W., 1999, J. Phys. Chem. A, 103**,** 5699

Blitz M.A., Johnson D.G., Pesa M., Pilling M.J., Robertson S.H., P W. Seakins, 1997, J. Chem. Soc. Faraday Trans., 93**,** 1473

Blitz M.A., Talbi D., Seakins P.W., Smith I.W.M., 2012, J. Phys. Chem. A, 116**,** 5877
Boggio-Pasqua M., Voronin A.I., Halvick P., Rayez J.C., 2000, Phys. Chem. Chem. Phys., 2**,** 1693
Boullart W., Devriendt K., Borms R., Peteers J., 1996, J. Phys. Chem., 100**,** 998
Brown R.D., Godfrey P.D., Cragg D.M., Rice E.H.N., Irvine W., M., Friberg P., Suzuki H., Ohishi M., Kaifu N., Morimoto M., 1985, ApJ, 297**,** 302
Brownsword R.A., Gatenby S.D., Herbert L.B., Smith I.W.M., Stewart D.W.A., Symonds A.C., 1996, J. Chem. Soc. Faraday Trans.**,** 7234
Brunken S., Gupta H., Gottlieb C.A., McCarthy M.C., Thaddeus P., 2007, ApJ, 664**,** L43
Cami J., Bernard-Salas J., Peeters E., Malek S.E., 2010, Science, 329**,** 1180
Canosa A., Sims I.R., Travers D., Smith I.W.M., Rowe B.R., 1997, A&A, 323**,** 644
Cernicharo J., Gottlieb C.A., Guélin M., Killian T.C., Paubert G., Thaddeus P., Vrtilek J.M., 1991, ApJ, 368**,** L39
Chabot M., Tuna T., Béroff K., Pino T., Padellec A.L., Désequelles P., Martinet G., Nguyen-Thi V.O., Carpentier Y., Petit F.L., Roueff E., Wakelam V., 2010, A&A A39
Chastaing D., James P.L., Sims I.R., Smith I.W.M., 1998, Faraday Discusion, 109**,** 165
Chastaing D., James P.L., Sims I.R., Smith I.W.M., 1999, Phys. Chem. Chem. Phys., 1**,** 2247
Chastaing D., Le Picard S.D., Sims I.R., Smith I.W.M., 2001, A&A, 365**,** 241
Chastaing D., Le Picard S.D., Sims I.R., Smith I.W.M., Geppert W.D., Naulin C., Costes M., 2000, Chem. Phys. Lett., 331**,** 170
Clary D.C., Haider N., Husain D., Kabir M., 1994, ApJ., 422**,** 416
Cordiner M.A., Buckle J.V., Wirström E.S., Olofsson A.O.H., Charnley S.B., 2013, ApJ, 770**,** 48
Costes M., Halvick P., Hickson K.M., Daugey N., Naulin C., 2009, ApJ, 703**,** 1179
Crutcher R.M., Churchwell E., Ziurys L.M., 1984, ApJ, 283**,** 668
Daranlot J., Hincelin U., Bergeat A., Costes M., Loison J.-C., Wakelam V., Hickson K.M., 2012, PNAS, 109**,** 10233
Daranlot J., Hu X., Xie C., Loison J.-C., Caubet P., Costes M., Wakelam V., Xie D., Guo H., Hickson K.M., 2013, Phys. Chem. Chem. Phys., 15**,** 13888
Daranlot J., Jorfi M., Xie C., Bergeat A., Costes M., Caubet P., Xie D., Guo H., Honvault P., Hickson K.M., 2011, Science, 334**,** 1538
Daugey N., Caubet P., Bergeat A., Costes M., Hickson K.M., 2008, Phys. Chem. Chem. Phys., 10**,** 729
Daugey N., Caubet P., Retail B., Costes M., Bergeat A., Dorthe G., 2005, PCCP, 7**,** 2921
Devriendt K., Van Look H., Ceursters B., Peeters J., 1996, Chem. Phys. Lett., 261**,** 450
Diaz-Tendero S., Sanchez G., Alcami M., Martin F., Hervieux P.A., Chabot M., Martinet G., Désesquelles P., Mezdari F., Wohrer-Béroff K., Negra S.D., Hamrita H., LePadellec A., Montagnon L., 2006, Int. J. Mass Spectrom., 252**,** 126
Dickens J.E., Irvine W.M., Snell R.L., Bergin E.A., Schloerb F.P., Pratap P., Miralles M.P., 2000, ApJ, 542**,** 870
Dunning Jr. T.H., 1989, J. Chem. Phys., 90**,** 1007
Florescu-Mitchell A.I., Mitchell J.B.A., 2006, Phys. Rep., 430**,** 277

Fossé D., Cernicharo J., Gerin M., Pierre C., 2001, ApJ, 552, 168
Galland N., Caralp F., Hannachi Y., Bergeat A., Loison J.-C., 2003, J. Phys. Chem. A, 107, 5419
Galland N., Caralp F., Rayez M.T., Hannachi Y., Loison J.C., Dorthe G., Bergeat A., 2001, Journal of Physical Chemistry A, 105, 9893
Gannon K.L., Glowacki D.R., Blitz M.A., Hughes K.J., Pilling M.J., Seakins P.W., 2007, J. Phys. Chem A., 111, 6679
Garrod R.T., Wakelam V., Herbst E., 2007, A&A, 467, 1103
Georgievskii Y., Klippenstein S.J., 2005, J. Chem. Phys., 122, 194103
Georgievskii Y., Klippenstein S.J., 2007, J. Phys. Chem A., 111, 3802
Georgievskii Y., Klippenstein S.J., 2011, IAU Symp., 7, 372
Georgievskii Y., Miller J.A., Klippenstein S.J., 2007, Phys. Chem. Chem. Phys., 9, 4259
Geppert W.D., Reignier D., Stoecklin T., Naulin C., Costes M., Chastaing D., Le Picard S.D., Sims I.R., Smith I.W.M., 2000, Phys. Chem. Chem. Phys., 2, 2873
Gerin M., Viala Y., Casoli F., 1993, A&A, 268, 212
Gerin M., Viala Y., Pauzat F., Ellinger Y., 1992, Astron. Astrophys., 266, 463
Giles K., Adams N.G., Smith D., 1989, Int. J. Mass Spectrom. Ion Proc., 89, 303
Goldsmith P.F., Li D., Krco M., 2007, ApJ, 654, 273
Goulay F., Soorkia S., Meloni G., Osborn D.L., Taatjes C.A., Leone S.R., 2011, Phys. Chem. Chem. Phys., 13, 20820
Goulay F., Trevitt A.J., Meloni G., Selby T.M., Osborn D.L., Taatjes C.A., Vereecken L., Leone S.R., 2009, J. Am. Chem. Soc., 131, 993
Gu X., Guo Y., Mebel A.M., Kaiser R.I., 2007, Chem. Phys. Lett., 449, 44
Guélin M., Neininger N., Cernicharo J., 1998, A&A, 335, L1
Guo Y., Gu X., Zhang F., Mebel A.M., Kaiser R.I., 2007, Phys. Chem. Chem. Phys., 9, 1972
Gupta H., Gottlieb C.A., McCarthy M.C., Thaddeus P., 2009, ApJ, 691, 1494
Haider N., Husain D., 1992, Z. Phys. Chem., 176, 133
Haider N., Husain D., 1993a, J. Chem. Soc. Faraday Trans., 89, 7
Haider N., Husain D., 1993b, J. Photochem. Photobiol. A, 70, 119
Harding L.B., Klippenstein S.J., 1998, Symp. (Int.) Comb., 27, 151
Harju J., Winnberg A., Wouterloot J.G.A., 2000, A&A, 353, 1065
Harland P.W., McIntosh B.J., 1985, Int. J. Mass Spectrom. Ion Proc., 67, 29
Hasegawa T.I., Herbst E., Leung C.M., 1992, ApJ, Supp. Series, 82, 167
Herbst E., 1994, Chemical Physics Letters, 222, 297
Herbst E., Klemperer W., 1973, Astrophys. J., 185, 505
Herbst E., Osamura Y., 2008, ApJ, 679, 1670
Herron J.T., Huie R.E., 1968, J. Phys. Chem., 72, 2538

Hersant F., Wakelam V., Dutrey A., Guilloteau S., Herbst E., 2009, A&A, 493, L49
Hincelin U., Wakelam V., Hersant F., Guilloteau S., Loison J.C., Honvault P., Troe J., 2011, A&A, 530, 61
Hobbs L.M., Black J.H., van Dishoeck E., 1983, ApJ, 271, L95
Hollenbach D., Kaufman M.J., Bergin E.A., Melnick G.J., 2009, ApJ, 690, 1497
Holmes J.L., Lossing F.P., Mayer P.M., 1993, Chem. Phys. Lett., 212, 134
Hunter E.P.L., Lias S.G., 1998, J. Phys. Chem. Ref. Data, 27, 413
Husain D., Ioannou A.X., 1997, J. Chem. Soc. Faraday Trans., 93, 3625
Husain D., Kirsch L.J., 1971a, Chem. Phys. Lett., 8, 543
Husain D., Kirsch L.J., 1971b, J. Chem. Soc. Faraday Trans., 67, 2025
Irvine W.M., Hoglund B., Friberg P., Askne J., Ellder J., 1981, ApJ, 248, L113
Joiner A., Mohr R.H., Yukich J.N., 2011, Phys. Rev. A, 035401
Ju L.-P., Xie T.-X., Zhang X., Han K.-L., 2005, Chem. Phys. Lett., 409, 249
Klippenstein S.J., Georgievskii Y., Harding L.B., 2006, Phys. Chem. Chem. Phys., 8, 1133
Knight J.S., Petrie S.A.H., Freeman C.G., McEwan M.J., McLean A.D., DeFrees D.J., 1988, JACS, 110, 5286
Langer W.D., Velusamy T., Kuiper T.B.H., Peng R., McCarthy M.C., Travers M.J., Kovacs A., Gottlieb C.A., Thaddeus P., 1997, ApJL, 480, L63
Li H.Y., Cheng W.C., Liu Y.L., Sun B.J., Huang C.Y., Chen K.T., Tang M.S., Kaiser R.I., Chang A.H.H., 2006, J. Chem. Phys., 124, 044307
Li J.-l., Huang X.-r., Bai H.-t., Geng C.-y., Yu G.-t., Sun C.-c., 2005, J. Mol. Struct. (THEOCHEM), 730, 207
Lias S.G., Bartmess J.E., Liebmann J.F., Holmes J.L., Levine R.D., Mallard W.G., 1988, J. Phys. Chem. Ref. Data, 17, suppl. No 1, 1
Loison J.-C., Bergeat A., 2009, Phys. Chem. Chem. Phys., 11, 655
Loison J.-C., Halvick P., Bergeat A., Hickson K.M., Wakelam V., 2012, MNRAS, 421, 1476
Loison J.C., Bergeat A., 2004, Phys. Chem. Chem. Phys., 6, 5396
Loison J.C., Bergeat A., Caralp F., Hannachi Y., 2006, J. Phys. Chem. A, 110, 13500
Martin J.M.L., Taylor P.R., 1995, J. Chem. Phys., 102, 8270
McElvany S.W., Dunlap B.I., O'Keefe A., 1987, J. Chem. Phys., 86, 715
McEwan M.J., Scott G.B.I., Adams N.G., Babcock L.M., Terzieva R., Herbst E., 1999, ApJ, 513, 287
Meads R.F., Maclagan R.G.A.R., Phillips L.F., 1993, J. Phys. Chem., 97, 3257
Mebel A.M., Kim G.S., Kislov V.V., Kaiser R.I., 2007a, J. Phys. Chem. A, 111, 6704
Mebel A.M., Kislov V.V., Hayashi M., 2007b, J. Chem. Phys., 126, 204310
Messing I., Carrington T., Filseth S.V., Sadowski C.M., 1980, Chem. Phys. Lett., 74, 56
Messing I., Filseth S.V., Sadowski C.M., Carrington T., 1981, J. Chem. Phys., 74, 3874
Mitchell M.B., Nava D.F., Stief L.J., 1986, J. Chem. Phys., 85, 3300

Monks P.S., Nesbitt F.L., Payne W.A., Scanlon M., Stief L.J., Shallcross D.E., 1995, J. Phys. Chem. A., 99**,** 17151

Nelson H.H., Helvajian H., Pasternack L., McDonald J.R., 1982, Chem. Phys., 73**,** 431

Nelson H.H., Pasternack L., Eyler J.R., McDonald J.R., 1981, Chem. Phys., 60**,** 231

Neufeld D.A., Wolfire M.G., Schilke P., 2005, ApJ, 628**,** 260

Ohishi M., Irvine W.M., Kaifu N., 1992, IAU Symp.**,** 171

Ohishi M., Kaifu N., 1998, Faraday Discuss., 109**,** 205

Pagani L., Olofsson A.O.H., Bergman P., Bernath P., Black J.H., Booth R.S., Buat V., Crovisier J., Curry C.L., Encrenaz P.J., Falgarone E., Feldman P.A., Fich M., Floren H.G., Frisk U., Gerin M., Gregersen E.M., Harju J., Hasegawa T., Hjalmarson Ã., Johansson L.E.B., Kwok S., Larsson B., Lecacheux A., Liljestram T., Lindqvist M., Liseau R., Mattila K., Mitchell G.F., Nordh L.H., Olberg M., Olofsson G., Ristorcelli I., Sandqvist A., Scheele F.v., Serra G., Tothill N.F., Volk K., Wiklind T., Wilson C.D., 2003, A&A, 402**,** L77

Paramo A., Canosa A., Le Picard S., D. b., Sims I.R., 2008, J. Phys. Chem. A,

Pedersen J.O.P., Opansky B.J., Leone S.R., 1993, J. Phys. Chem., 97**,** 6822

Phillips L.F., 1990, Chem. Phys. Lett., 165**,** 545

Plessis S., Carrasco N., Dobrijevic M., Pernot P., 2012, Icarus, 219**,** 254

Prasad S.S., Trafdar S.P., 1983, ApJ, 267**,** 603

Pratap P., Dickens J.E., Snell R.L., Miralles M.P., Bergin E.A., Irvine W.M., Schloerb F.P., Charnley S.B., 1997, ApJ, 486**,** 862

Reisler H., Mangir M., Wittig C., 1980a, Chem. Phys., 47**,** 49

Reisler H., Mangir M.S., Wittig C., 1980b, J. Chem. Phys., 73**,** 2280

Sabbah H., Biennier L., Sims I.R., Georgievskii Y., Klippenstein S.J., Smith I.W.M., 2007, Science, 317**,** 102

Sakai N., Saruwatari O., Sakai T., Takano S., Yamamoto S., 2010, A&A, 512**,** A31

Sato S., Sugawara K., Ishikawa Y., 1979, Chem. Phys. Lett., 68**,** 557

Savic I., Gerlich D., 2005, Phys. Chem. Chem. Phys., 7**,** 1026

Schacke H., Schmatjko K.J., Wolfrum. J., 1973, Ber Bunsenges. Physik. Chem., 77**,** 248

Schmatjko K.J., Wolfrum J., 1977, Int. Symp. Comb., 16**,** 819

Schmatjko K.J., Wolfrum J., 1978, Ber. Bunsen-Ges. Phys. Chem., 82**,** 419

Scott G.B.I., Fairley D.A., Freeman C.G., McEwan M.J., Anicich V.G., 1999, J. Phys. Chem. A, 103**,** 1073

Semenov D., Hersant F., Wakelam V., Dutrey A., Chapillon E., Guilloteau S., Henning T., Launhardt R., Pitu V., Schreyer K., 2010, A&A, 522**,** A42

Sims I.R., Queffelec J.-L., Travers D., Rowe B.R., Herbert L.B., Karthäuser J., Smith I.W.M., 1993, Chem. Phys. Lett., 211**,** 461

Smith I.W.M., 2006, Angewandte Chemie - International Edition, 45**,** 2842

Smith I.W.M., 2011, ARA&A, 49**,** 29

Smith I.W.M., Sage A.M., Donahue N.M., Herbst E., Park I.H., 2006, Faraday Discuss., 133**,** 1

Suutarinen A., Geppert W.D., Harju J., Heikkila A., Hotzel S., Juvela M., Millar T.J., Walsh C., Wouterloot J.G.A., 2011, A&A, 531**,** A121
Takahashi J., Yamashita K., 1996, J. Chem. Phys., 104**,** 6613
Takano S., Masuda A., Hirahara Y., Suzuki H., Ohishi M., Ishikawa S.-I., Kaifu N., Kasai Y., Kawaguchi K., Wilson T.L., 1998, A&A, 329**,** 1156
Thaddeus P., Gottlieb C.A., Gupta H., Brünken S., McCarthy M.C., Agundez M., Guélin M., Cernicharo J., 2008, ApJ, 677**,** 1132
Titarchuk T.A., Halpern J.B., 1995, Chem. Phys. Lett., 232**,** 192
Turner B.E., E. H., Terzieva R., 2000, ApJ Supp. Series, 126**,** 427
Turner B.E., Terzieva R., Herbst E., 1999, ApJ, 518**,** 699
Van Orden A., Saykally R.J., 1998, Chem. Rev., 98**,** 2313
Wakelam V., Herbst E., Loison J.C., Smith I.W.M., Chandrasekaran V., Pavone B., Adams N.G., Bacchus-Montabonel M.C., Bergeat A., Béroff K., Bierbaum V.M., Chabot M., Dalgarno A., Van Dishoeck E.F., Faure A., Geppert W.D., Gerlich D., Galli D., Hébrard E., Hersant F., Hickson K.M., Honvault P., Klippenstein S.J., Le Picard S., Nyman G., Pernot P., Schlemmer S., Selsis F., Sims I.R., Talbi D., Tennyson J., Troe J., Wester R., Wiesenfeld L., 2012, ApJ, Supp. Series, 199
Wakelam V., Loison J.C., Herbst E., Talbi D., Quan D., Caralp F., 2009, A&A, 495**,** 513
Wakelam V., Smith I.W.M., Herbst E., Troe J., Geppert W., Linnartz H., Oberg K., Roueff E., Agundez M., Pernot P., Cuppen H.M., Loison J.C., Talbi D., 2010, Space Sciences Rev., 156**,** 13
Wang J., Ding Y.-h., Sun C.-c., 2005, ChemPhysChem, 6**,** 431
Wang J., Ding Y.-h., Sun C.-c., 2006, ChemPhysChem, 7**,** 710
Whyte A.R., Phillips L.F., 1983, Chem. Phys. Lett., 98**,** 590
Woodall J., Agundez M., Markwick-Kemper A.J., Millar T.J., 2007, A&A, 466**,** 1197
Woon D.E., 2006, Chem. Phys., 331**,** 67
Woon D.E., Herbst E., 2009, ApJ, Supp. Series, 185**,** 273
Zhao X.-l., Zhang J.-x., Liu J.-y., Li X.-t., Li Z.-s., 2007, Chem. Phys. Lett., 436**,** 41
Zhao Y., Truhlar D., 2008, Theor. Chem. Acc., 120**,** 215
Zhu Z., Zhang Z., Huang C., Pei L., Chen C., Chen Y., 2003, J. Phys. Chem. A., 107**,** 10288

**Table 1: Summary of reactions review. (Full Table)**

For carbon atoms reactions T = 10 K only except for C + $C_3$ and C + $C_2H_2$.

| | Reaction | ΔE kJ/mol | α | β | γ | $F_0$ | g | ref |
|---|---|---|---|---|---|---|---|---|
| 13. | $^3C + {}^1C_3 \rightarrow {}^3C_4 + h\nu$ | -495 | 4.0e-14 | -1.0 | 0 | 3 | 0 | (Wakelam *et al.* 2009) |
| 14. | $^3C + {}^3C_4 \rightarrow {}^{1,3}C_2 + {}^1C_3$ | -101 | 2.4e-10 | | | 3 | 0 | (Wakelam et al. 2009) |
| 15. | $^3C + {}^1C_5 \rightarrow {}^1C_3 + {}^1C_3$ | -138 | 1.5e-10 | | | 3 | 0 | We consider an exit Transition State below the entrance channel (Nelson *et al.* 1982) and we assume that k = $0.5 \times k_{capture}$ due to back dissociation as it needs triplet-singlet crossing, the $^1C_3$ + $^3C_3$ channel being endothermic by 64 kJ/mol. |
| 16. | $^3C + {}^3C_6 \rightarrow {}^{3,1}C_4 + {}^1C_3$<br>$\rightarrow {}^1C_5 + {}^{1,3}C_2$ | -147<br>-110 | 2.0e-10<br>4.0e-11 | | | 3<br>3 | 0<br>0 | (Diaz-Tendero *et al.* 2006, Wakelam et al. 2009) |
| 17. | $^3C + {}^1C_7 \rightarrow {}^1C_3 + {}^1C_5$ | -131 | 2.0e-10 | | | 3 | 0 | We consider an exit Transition State below the entrance channel (Nelson *et al.* 1982) and we assume that k = $0.8 \times k_{capture}$. |
| 18. | $^3C + {}^3C_8 \rightarrow {}^1C_7 + {}^{1,3}C_2$<br>$\rightarrow {}^{3,1}C_6 + {}^1C_3$<br>$\rightarrow {}^1C_5 + {}^3C_4$ | -69<br>-105<br>-80 | 1.0e-11<br>1.6e-10<br>7.0e-11 | | | 3<br>3<br>3 | 0<br>0<br>0 | |
| 19. | $^3C + {}^1C_9 \rightarrow {}^1C_7 + {}^1C_3$<br>$\rightarrow {}^1C_5 + {}^1C_5$ | -116<br>-116 | 2.1e-10<br>0.3e-10 | | | 3<br>3 | 0<br>0 | (Diaz-Tendero et al. 2006, Wakelam et al. 2009). This reaction occurs through triplet-singlet crossing but it involves a large system so the adduct lifetime is long. |
| 20. | $^3C + {}^3C_{10} \rightarrow {}^{3,1}C_8 + {}^1C_3$<br>$\rightarrow {}^{3,1}C_6 + {}^1C_5$ | -108<br>-90 | 1.8e-10<br>6.0e-11 | | | 3<br>3 | 0<br>0 | The most stable form of $C_{10}$ is the cyclic (c-$^1C_{10}$) one (Van Orden & Saykally 1998) |
| 21. | $^3C + {}^1C_{11} \rightarrow {}^1C_9 + {}^1C_3$<br>$\rightarrow {}^1C_7 + {}^1C_5$ | | 2.1e-10<br>0.3e-10 | | | 3<br>3 | 0<br>0 | Same as C + $C_9$ |
| | | | | | | | | |
| 22. | $^3C + CH(X^2\Pi) \rightarrow {}^{1,3}C_2 + {}^2H$ | -257 | 2.4e-10 | | | 2 | 0 | (Boggio-Pasqua *et al.* 2000) |
| 23. | $^3C + C_2H(X^2\Sigma) \rightarrow {}^1C_3 + {}^2H$ | -235 | 2.4e-10 | | | 3 | 0 | Same as C + CH |
| 24. | $^3C$ + c-$^2C_3H(X^2A')$ $\rightarrow {}^3C_4 + {}^2H$ | -175 | 2.4e-10 | | | 3 | 0 | Same as C + CH |
| | $\rightarrow {}^{1}C_2 + {}^{2}C_2H$ | -41 | 0 | | | 0 | 0 | |
| 25. | ${}^{3}C$ + l-${}^{2}C_3H(X^2A')$ $\rightarrow {}^{3}C_4 + {}^{2}H$<br>$\rightarrow {}^{1}C_2 + {}^{2}C_2H$ | -163<br>-29 | 2.4e-10<br>0 | | | 3<br>0 | 0<br>0 | Same as C + CH |
| 26. | ${}^{3}C + {}^{2}C_4H$ $\rightarrow {}^{1}C_5 + {}^{2}H$<br>$\rightarrow {}^{1}C_3 + {}^{2}C_2H$ | -233<br>-136 | 2.4e-10<br>0 | | | 3<br>0 | 0<br>0 | Same as C + CH |
| 27. | ${}^{3}C + {}^{2}C_5H$ $\rightarrow {}^{3}C_6 + {}^{2}H$ | -117 | 2.4e-10 | | | 3 | 0 | Same as C + CH |
| 28. | ${}^{3}C + {}^{2}C_6H$ $\rightarrow {}^{1}C_7 + {}^{2}H$<br>$\rightarrow {}^{1}C_3 + {}^{2}C_4H$ | -184<br>-82 | 2.4e-10<br>0 | | | 3<br>0 | 0<br>0 | |
| 29. | ${}^{3}C + {}^{2}C_7H$ $\rightarrow {}^{3}C_8 + {}^{2}H$ | | 2.4e-10 | | | 3 | 0 | Same as C + CH |
| 30. | ${}^{3}C + {}^{2}C_8H$ $\rightarrow {}^{1}C_9 + {}^{2}H$ | -165 | 2.4e-10 | | | 3 | 0 | Same as C + CH |
| 31. | ${}^{3}C + {}^{2}C_9H$ $\rightarrow {}^{3}C_{10} + {}^{2}H$ | | 2.4e-10 | | | 3 | 0 | Same as C + CH. It should be noted that the most stable form of $C_{10}$ is the cyclic (c-${}^{1}C_{10}$) one (Van Orden & Saykally 1998) |
| 32. | ${}^{3}C + {}^{2}C_{10}H$ $\rightarrow {}^{1}C_{11} + {}^{2}H$ | | 2.4e-10 | | | 3 | 0 | Same as C + CH |
| | | | | | | | | |
| 33. | ${}^{3}C + {}^{1}C_2H_2$ $\rightarrow {}^{1}C_3 + {}^{1}H_2$<br>$\rightarrow$ c-${}^{2}C_3H + {}^{2}H$<br>$\rightarrow$ l-${}^{2}C_3H + {}^{2}H$ | -115<br>-12<br>+1±4 | 2.6e-10<br>4.1e-11<br>7.8e-12 | -0.07<br>-0.39<br>1.08 | 0<br>2<br>0 | 1.2<br>1.4<br>1.8 | 0.9<br>0.8<br>2 | (Costes *et al.* 2009, Chastaing *et al.* 2001) |
| 34. | ${}^{3}C + {}^{1}C_4H_2$ $\rightarrow {}^{2}C_5H + {}^{2}H$<br>$\rightarrow {}^{1}C_5 + {}^{1}H_2$<br>$\rightarrow {}^{1}C_3 + {}^{1}C_2H_2$ | -36<br>-103<br>-128 | 3.0e-10<br>0<br>0 | | | 2 | 0 | See text. (Gu *et al.* 2007) |
| 35. | ${}^{3}C + {}^{1}C_5H_2$ $\rightarrow {}^{2}C_6H + {}^{2}H$<br>$\rightarrow {}^{3}C_6 + {}^{1}H_2$ | -265<br>-211 | 3.0e-10<br>0 | | | 2<br>0 | 0<br>0 | See text. We consider a high exit barrier for $H_2$ formation |
| 36. | ${}^{3}C + {}^{1}C_6H_2$ $\rightarrow {}^{2}C_7H + {}^{2}H$ | | 3.0e-10 | | | 2 | 0 | See text. |
| 37. | ${}^{3}C + {}^{1}C_7H_2$ $\rightarrow {}^{2}C_8H + {}^{2}H$ | | 3.0e-10 | | | 2 | 0 | See text. |
| 38. | ${}^{3}C + {}^{1}C_8H_2$ $\rightarrow {}^{2}C_9H + {}^{2}H$ | | 3.0e-10 | | | 2 | 0 | See text. |
| 39. | ${}^{3}C + {}^{1}C_9H_2$ $\rightarrow {}^{2}C_{10}H + {}^{2}H$ | | 3.0e-10 | | | 2 | 0 | See text. |
| 40. | ${}^{3}C + {}^{1}C_{10}H_2$ $\rightarrow {}^{1}C_3 + {}^{1}C_8H_2$<br>$\rightarrow {}^{1}C_5 + {}^{1}C_6H_2$<br>$\rightarrow {}^{1}C_7 + {}^{1}C_4H_2$<br>$\rightarrow {}^{1}C_9 + {}^{1}C_2H_2$<br>$\rightarrow {}^{2}C_2H + {}^{2}C_9H$<br>$\rightarrow {}^{2}C_3H + {}^{2}C_8H$ | | 4.0e-11<br>4.0e-11<br>4.0e-11<br>4.0e-11<br>4.0e-11<br>4.0e-11 | | | 2<br>2<br>2<br>2<br>2<br>2 | 0<br>0<br>0<br>0<br>0<br>0 | The products are likely to be $C_{11}H$ + H but $C_{11}H$ is not present in the model. We distribute the products over various spin forbidden and slightly endothermic exit channels to avoid the creation of an artificial sink. |
| | → $^2C_4H + ^2C_7H$<br>→ $^2C_5H + ^2C_6H$ | | 4.0e-11<br>4.0e-11 | | | 2<br>2 | 0<br>0 | |
| | | | | | | | | |
| 41. | $^3C + CCN(X^2\Pi)$ → $^2CN + ^{1,3}C_2$<br>→ $^1C_3 + ^4N$ | -133<br>-103 | 2.4e-10<br>0 | | | 3<br>0 | 0<br>0 | See text. |
| 42. | $^3C + C_3N(X^2\Sigma^+)$ → $^2CN + ^1C_3$ | -142 | 2.4e-10 | | | 3 | 0 | See text. |
| 43. | $^3C + C_4N(X^2\Pi)$ → $^2C_3N + ^{1,3}C_2$<br>→ $^1C_3 + ^2C_2N$<br>→ $^2CN + ^3C_4$ | -35<br>-44<br>-47 | 8.0e-11<br>8.0e-11<br>8.0e-11 | | | 3<br>3<br>3 | 0<br>0<br>0 | See text. We consider no barrier to the exit channels even for $^1C_3 + ^2C_2N$ by comparison with $^1C_2H_2 + ^2C_2N$ reaction (Wang *et al.* 2006) |
| 44. | $^3C + C_5N(X^2\Sigma^+)$ → $^2C_3N + ^1C_3$<br>→ $^2CN + ^1C_5$ | -157<br>-161 | 1.2e-10<br>1.2e-10 | | | 3<br>3 | 0<br>0 | See text. We consider no exit barrier for the $C_3N + C_3$ and $CN + C_5$ exit channels by comparison with $CN + C_2H_4$ (Sims et al. 1993, Gannon et al. 2007) and $C_3N$ reactivity (Petrie & Osamura 2004) |
| 45. | $^3C + C_6N(X^2\Pi)$ → $^2C_4N + ^1C_3$<br>→ $^2C_3N + ^3C_4$<br>→ $^2CN + ^3C_6$ | -79<br>-13<br>-8 | 1.6e-10<br>6.0e-11<br>2.0e-11 | | | 3<br>3<br>3 | 0<br>0<br>0 | See text. We consider no exit barrier for the $^2C_3N + ^3C_4$ and $^2CN + ^3C_6$ exit channels but there is likely to be one for $^2C_4N + ^1C_3$ formation. The branching ratio is roughly estimated from the exothermicities. |
| 46. | $^3C + C_7N(X^2\Sigma^+)$ → $^2C_5N + ^1C_3$<br>→ $^2CN + ^1C_7$ | -87<br>-117 | 1.2e-10<br>1.2e-10 | | | 3<br>3 | 0<br>0 | Simplified exit channels |
| 47. | $^3C + C_8N(X^2\Pi)$ → $^2C_6N + ^1C_3$<br>→ $^2C_4N + ^1C_5$<br>→ $^2C_3N + ^3C_6$<br>→ $^2CN + ^3C_8$ | -115<br>-56<br>-25<br>-66 | 1.2e-10<br>4.0e-11<br>4.0e-11<br>4.0e-11 | | | 3<br>3<br>3<br>3 | 0<br>0<br>0<br>0 | Branching ratio roughly estimated from exothermicities. |
| 48. | $^3C + C_9N(X^2\Sigma^+)$ → $^2C_7N + ^1C_3$<br>→ $^2CN + ^1C_9$ | | 1.2e-10<br>1.2e-10 | | | 3<br>3 | 0<br>0 | Simplified exit channels |
| 49. | $^3C + C_{10}N(X^2\Pi)$ → $^2C_8N + ^1C_3$<br>→ $^2CN + ^3C_{10}$ | | 1.2e-10<br>1.2e-10 | | | 3<br>3 | 0<br>0 | Simplified exit channels |
| | | | | | | | | |
| 50. | $^3C + ^1HC_3N$ → $^2C_4N + ^2H$ | -19 | 1.0e-10 | | | 3 | 0 | (Li et al. 2006) |
| 51. | $^3C + ^1HC_5N$ → $^2C_6N + ^2H$ | -36 | 1.0e-10 | | | 3 | 0 | Equal to $C + HC_3N$, thermochemistry calculated at the M06-2X/cc-pVTZ level using Gaussian09. |
| **T** | $^3C + ^1HC_7N$ → $^2C_8N + ^2H$ | -49 | 1.0e-10 | | | 3 | 0 | Equal to $C + HC_3N$, thermochemistry calculated at the M06-2X/cc-pVTZ level using Gaussian09 |
| 52. | $^3C + {}^1HC_9N \rightarrow {}^2C_{10}N + {}^2H$ | | 1.0e-10 | | | 3 | 0 | Equal to C + $HC_3N$. |
| | | | | | | | | |
| 53. | $^3C + {}^1C_3O \rightarrow {}^3C_3 + {}^1CO$<br>$\rightarrow {}^1C_3 + {}^1CO$ | -126<br>-328 | 3.0e-10<br>0 | | | 3<br>0 | 0<br>0 | See text. |
| 54. | $^3C + {}^1C_5O \rightarrow {}^3C_5 + {}^1CO$<br>$\rightarrow {}^1C_5 + {}^1CO$ | <br>-387 | 3.0e-10<br>0 | | | 3<br>0 | 0<br>0 | See text. |
| 55. | $^3C + {}^1C_7O \rightarrow {}^{1,3}C_7 + {}^1CO$ | | 3.0e-10 | | | 3 | 0 | See text. |
| 56. | $^3C + {}^1C_9O \rightarrow {}^{1,3}C_9 + {}^1CO$ | | 3.0e-10 | | | 3 | 0 | See text. |
| | | | | | | | | |
| 57. | $^4N + {}^1C_2 \rightarrow {}^3C + {}^2CN$ | -159 | 2.0e-10 | 0.17 | 0 | 2 | 0 | See text. |
| 58. | $^4N + {}^3C_4 \rightarrow {}^1C_3 + {}^2CN$ | -260 | 3.0e-11 | 0.17 | 0 | 3 | 0 | Comparison with the N + $^3C_2$ reaction (Becker et al. 2000) |
| 59. | $^4N + {}^3C_6 \rightarrow {}^1C_5 + {}^2CN$<br>$\rightarrow {}^1C_3 + {}^2C_3N$ | -269<br>-265 | 1.5e-11<br>1.5e-11 | 0.17<br>0.17 | 0<br>0 | 3<br>3 | 0<br>0 | Comparison with the N + $^3C_2$ reaction (Becker et al. 2000). Branching ratio roughly estimated from the exothermicities. |
| 60. | $^4N + {}^3C_8 \rightarrow {}^1C_7 + {}^2CN$<br>$\rightarrow {}^1C_5 + {}^2C_3N$<br>$\rightarrow {}^1C_3 + {}^2C_5N$ | | 1.0e-11<br>1.0e-11<br>1.0e-11 | 0.17<br>0.17<br>0.17 | 0<br>0<br>0 | 3<br>3<br>3 | 0<br>0<br>0 | By comparison with N + $^3C_2$ reaction (Becker et al. 2000). Branching ratio roughly estimated with exothermicities. |
| 61. | $^4N + {}^3C_{10} \rightarrow {}^1C_9 + {}^2CN$<br>$\rightarrow {}^1C_7 + {}^2C_3N$<br>$\rightarrow {}^1C_5 + {}^2C_5N$<br>$\rightarrow {}^1C_3 + {}^2C_7N$ | | 1.0e-11<br>1.0e-11<br>1.0e-11<br>1.0e-11 | 0.17<br>0.17<br>0.17<br>0.17 | 0<br>0<br>0<br>0 | 3<br>3<br>3<br>3 | 0<br>0<br>0<br>0 | Comparison with the N + $^3C_2$ reaction (Becker et al. 2000). Branching ratio roughly estimated from the exothermicities. |
| 62. | $^4N + {}^1C_{3,5,7,9} \rightarrow {}^3C_{2,4,6,8} + {}^2CN$ | | 0 (10 K) | | | | | Ab-initio calculations for the $^4N + {}^1C_3$ reaction show a barrier. |
| | | | | | | | | |
| 63. | $^4N + {}^2CH \rightarrow {}^2CN + {}^2H$ | -416 | 1.6e-10 | 0.17 | 0 | 1.6 | 7 | Deduced from (Brownsword et al. 1996) |
| 64. | $^4N + {}^2C_2H \rightarrow {}^2CCN + {}^2H$<br>$\rightarrow {}^1HCN + {}^3C$<br>$\rightarrow {}^1HNC + {}^3C$ | -132<br>-187<br>-132 | 1.5e-10<br>8.0e-12<br>2.0e-12 | 0.17<br>0.17<br>0.17 | 0<br>0<br>0 | 2<br>2<br>2 | 0<br>0<br>0 | Deduced from (Brownsword et al. 1996). Branching ratio from Ab-initio and RRKM calculations (this work). |
| 65. | $^4N + l\text{-}{}^2C_3H \rightarrow {}^2C_3N + {}^2H$<br>$\rightarrow {}^2CN + {}^2C_2H$<br>$\rightarrow {}^1HCN + {}^3C_2$ | -292<br>-212<br>-218 | 1.1e-10<br>5.0e-11<br>0 | 0.17<br>0.17<br>0 | 0<br>0<br>0 | 3<br>3<br>0 | 0<br>0<br>0 | See text. Branching ratio estimated from $HC_3N$ photodissociation (Silva *et al.* 2009). The $^1HCN + {}^3C_2$ exit channel is neglected as it needs a tight isomerization transition state. |

| | | | | | | | | | |
|---|---|---|---|---|---|---|---|---|---|
| 66. | $^{4}N + c\text{-}^{2}C_3H$ | → $^{2}C_3N + {}^{2}H$ | -280 | 1.1e-10 | 0.17 | 0 | 3 | 0 | See text. The c-$^{2}C_3H$ is assumed to have similar reactivity to l-$^{2}C_3H$. |
| | | → $^{2}CN + {}^{2}C_2H$ | -200 | 5.0e-11 | 0.17 | 0 | 3 | 0 | |
| | | → $^{1}HCN + {}^{3}C_2$ | -205 | 0 | 0 | 0 | 0 | 0 | |
| 67. | $^{4}N + {}^{2}C_4H$ | → $^{2}C_4N + {}^{2}H$ | -224 | 7.0e-11 | 0.17 | 0 | 3 | 0 | See text. The branching ratios are roughly estimated from exothermicities. |
| | | → $^{2}CN + {}^{2}C_3H$ | -125 | 1.0e-11 | 0.17 | 0 | 3 | 0 | |
| | | → $^{1}HC_3N + {}^{3}C$ | -217 | 7.0e-11 | 0.17 | 0 | 3 | 0 | |
| | | → $^{1}HCN + {}^{3}C_3$ | -121 | 1.0e-11 | 0.17 | 0 | 3 | 0 | |
| | | → $^{1}HNC + {}^{3}C_3$ | -66 | 0 | 0 | 0 | 0 | 0 | |
| 68. | $^{4}N + {}^{2}C_5H$ | → $^{2}C_5N + {}^{2}H$ | -225 | 6.0e-11 | 0.17 | 0 | 3 | 0 | See text. The branching ratios are roughly estimated from exothermicities. |
| | | → $^{2}C_3N + {}^{2}C_2H$ | -147 | 1.0e-11 | 0.17 | 0 | 3 | 0 | |
| | | → $^{2}CN + {}^{2}C_4H$ | -115 | 1.0e-11 | 0.17 | 0 | 3 | 0 | |
| | | → $^{1}HC_3N + {}^{3}C_2$ | -201 | 3.0e-11 | 0.17 | 0 | 3 | 0 | |
| | | → $^{1}HCN + {}^{3}C_4$ | -216 | 3.0e-11 | 0.17 | 0 | 3 | 0 | |
| | | → $^{1}HNC + {}^{3}C_4$ | -161 | 1.0e-11 | 0.17 | 0 | 3 | 0 | |
| 69. | $^{4}N + {}^{2}C_6H$ | → $^{2}C_6N + {}^{2}H$ | -227 | 1.0e-10 | 0.17 | 0 | 3 | 0 | See text. The branching ratios are roughly estimated from exothermicities. |
| | | → $^{1}HC_5N + {}^{3}C$ | -170 | 6.0e-11 | 0.17 | 0 | 3 | 0 | |
| 70. | $^{4}N + {}^{2}C_7H$ | → $^{2}C_7N + {}^{2}H$ | | 1.0e-10 | 0.17 | 0 | 3 | 0 | The branching ratios are roughly estimated from exothermicities. The production of both HCN and HNC are neglected. |
| | | → $^{2}C_5N + {}^{2}C_2H$ | | 2.0e-11 | 0.17 | 0 | 3 | 0 | |
| | | → $^{2}C_3N + {}^{2}C_4H$ | | 2.0e-11 | 0.17 | 0 | 3 | 0 | |
| | | → $^{2}CN + {}^{2}C_6H$ | | 2.0e-11 | 0.17 | 0 | 3 | 0 | |
| | | → $^{1}HCN + {}^{3}C_6$ | | 0 | 0 | 0 | 0 | 0 | |
| | | → $^{1}HNC + {}^{3}C_6$ | | 0 | 0 | 0 | 0 | 0 | |
| 71. | $^{4}N + {}^{2}C_8H$ | → $^{2}C_8N + {}^{2}H$ | | 8.0e-11 | 0.17 | 0 | 3 | 0 | See text. The branching ratios are roughly estimated from exothermicities. |
| | | → $^{1}HC_7N + {}^{3}C$ | | 8.0e-11 | 0.17 | 0 | 3 | 0 | |
| 72. | $^{4}N + {}^{2}C_9H$ | → $^{2}C_9N + {}^{2}H$ | | 1.2e-10 | 0.17 | 0 | 3 | 0 | The branching ratios are roughly estimated from exothermicities. The production of both HCN and HNC are neglected. |
| | | → $^{2}C_7N + {}^{2}C_2H$ | | 1.0e-11 | 0.17 | 0 | 3 | 0 | |
| | | → $^{2}C_5N + {}^{2}C_4H$ | | 1.0e-11 | 0.17 | 0 | 3 | 0 | |
| | | → $^{2}C_3N + {}^{2}C_6H$ | | 1.0e-11 | 0.17 | 0 | 3 | 0 | |
| | | → $^{2}CN + {}^{2}C_8H$ | | 1.0e-11 | 0.17 | 0 | 3 | 0 | |
| 73. | $^{4}N + {}^{2}C_{10}H$ | → $^{2}C_{10}N + {}^{2}H$ | | 8.0e-11 | 0.17 | 0 | 3 | 0 | See text. The branching ratios are roughly estimated from exothermicities. |
| | | → $^{1}HC_9N + {}^{3}C$ | | 8.0e-11 | 0.17 | 0 | 3 | 0 | |

| | | | | | | | | |
|---|---|---|---|---|---|---|---|---|
| 74. | $^{4}N + CN(X^{2}\Sigma^{+}) \rightarrow {}^{3}C + {}^{1}N_2$ | -191 | 9.0e-11 | 0.42 | 0 | 1.4 | 0 | (Daranlot et al. 2012) |
| 75. | $^{4}N + C_2N(X^{2}\Pi) \rightarrow {}^{2}CN + {}^{2}CN$<br>$\rightarrow {}^{3}C_2 + {}^{1}N_2$ | -21<br>-317 | 9.0e-11<br>0 | 0.17<br>0 | 0<br>0 | 3<br>3 | 0<br>0 | See text. |
| 76. | $^{4}N + {}^{2}C_3N(X^{2}\Sigma^{+}) \rightarrow {}^{2}CN + C_2N$<br>$\rightarrow {}^{1}C_3 + {}^{1}N_2$ | -39<br>-333 | 9.0e-11<br>0 | 0.17<br>0 | 0<br>0 | 3<br>3 | 0<br>0 | See text. |
| 77. | $^{4}N + {}^{2}C_4N(X^{2}\Pi) \rightarrow {}^{2}CN + {}^{2}C_3N$<br>$\rightarrow {}^{3}C_4 + {}^{1}N_2$ | -194<br>-267 | 9.0e-11<br>0 | 0.17<br>0 | 0<br>0 | 3<br>3 | 0<br>0 | See text. |
| 78. | $^{4}N + {}^{2}C_5N(X^{2}\Sigma^{+}) \rightarrow {}^{2}CN + {}^{2}C_4N$<br>$\rightarrow {}^{2}C_2N + {}^{2}C_3N$<br>$\rightarrow {}^{1}C_5 + {}^{1}N_2$ | -152<br>-54<br>-352 | 9.0e-11<br>0<br>0 | 0.17<br>0<br>0 | 0<br>0<br>0 | 3<br>3<br>0 | 0<br>0<br>0 | See text. |
| 79. | $^{4}N + {}^{2}C_6N(X^{2}\Pi) \rightarrow {}^{2}CN + {}^{2}C_5N$<br>$\rightarrow {}^{2}C_3N + {}^{2}C_3N$<br>$\rightarrow {}^{3}C_6 + {}^{1}N_2$ | -116<br>-131<br>-199 | 4.0e-11<br>5.0e-11<br>0 | 0.17<br>0.17<br>0 | 0<br>0<br>0 | 3<br>3<br>0 | 0<br>0<br>0 | See text. |
| 80. | $^{4}N + {}^{2}C_7N(X^{2}\Sigma^{+}) \rightarrow {}^{2}CN + {}^{2}C_6N$ | | 9.0e-11 | 0.17 | 0 | 3 | 0 | See text. |
| 81. | $^{4}N + {}^{2}C_8N(X^{2}\Pi) \rightarrow {}^{2}CN + {}^{2}C_7N$<br>$\rightarrow {}^{2}C_3N + {}^{2}C_5N$<br>$\rightarrow {}^{3}C_8 + {}^{1}N_2$ | | 4.0e-11<br>5.0e-11<br>0 | 0.17<br>0.17<br>0 | 0<br>0<br>0 | 3<br>3<br>0 | 0<br>0<br>0 | See text. |
| 82. | $^{4}N + {}^{2}C_9N(X^{2}\Sigma^{+}) \rightarrow {}^{2}CN + {}^{2}C_8N$ | | 9.0e-11 | 0.17 | 0 | 3 | 0 | See text. |
| 83. | $^{4}N + {}^{2}C_{10}N(X^{2}\Pi) \rightarrow {}^{2}CN + {}^{2}C_9N$<br>$\rightarrow {}^{2}C_3N + {}^{2}C_7N$<br>$\rightarrow {}^{2}C_5N + {}^{2}C_5N$<br>$\rightarrow {}^{3}C_{10} + {}^{1}N_2$ | | 3.0e-11<br>3.0e-11<br>3.0e-11<br>0 | 0.17<br>0.17<br>0.17<br>0 | 0<br>0<br>0<br>0 | 3<br>3<br>3<br>0 | 0<br>0<br>0<br>0 | See text. |
| 84. | | | | | | | | |
| 85. | $^{3}O + {}^{1}C_2 \rightarrow {}^{3}C + {}^{1}CO$ | -381 | 2.0e-10 | 0.17 | 0 | 3 | 0 | See text. |
| 86. | $^{3}O + {}^{1}C_3 \rightarrow {}^{3}C_2 + {}^{1}CO$ | -352 | 0 (10K) | | | | | We assume the presence of a barrier for this reaction. |
| 87. | $^{3}O + {}^{3}C_4 \rightarrow {}^{3}C_3 + {}^{1}CO$<br><br>$\rightarrow {}^{1}C_3 + {}^{1}CO$ | -378<br>-582 | 1.0e-10<br>0 | 0.17<br>0 | 0<br>0 | 2<br>0 | 0<br>0 | See text. |
| 88. | $^{3}O + {}^{1}C_5 \rightarrow {}^{3}C_4 + {}^{1}CO$ | -389 | 0 (10K) | | | | | We assume the presence of a barrier for this reaction. |
| 89. | $^{3}O + {}^{3}C_6 \rightarrow {}^{3}C_5 + {}^{1}CO$<br>$\rightarrow {}^{1}C_5 + {}^{1}CO$ | <br>-591 | 1.0e-10<br>0 | 0.17<br>0 | 0<br>0 | 2<br>0 | 0<br>0 | See text. |
| 90. | $^{3}O + {}^{1}C_7 \rightarrow {}^{3}C_6 + {}^{1}CO$ | -373 | 0 (10K) | | | | | We assume the presence of a barrier for this reaction. |
| 91. | $^3O + ^3C_8 \rightarrow ^3C_7 + ^1CO$<br>$\rightarrow ^1C_7 + ^1CO$ | <br>-550 | 1.0e-10<br>0 | 0.17<br>0 | 0<br>0 | 2<br>0 | 0<br>0 | See text. |
| 92. | $^3O + ^1C_9 \rightarrow ^3C_8 + ^1CO$ | -449 | 0 (10K) | | | | | We assume the presence of a barrier for this reaction. |
| 93. | $^3O + ^3C_{10} \rightarrow ^3C_9 + ^1CO$ | | 1.0e-10 | 0.17 | 0 | 2 | 0 | See text. |
| | | | | | | | | |
| 94. | $^3O + ^2CH(X^2\Pi) \rightarrow ^1CO + ^2H$<br><br>$\rightarrow ^3CO + ^2H$ | -736<br>-153 | 7.0e-11 | 0 | 0 | 2 | 0 | (Messing *et al.* 1980, Messing *et al.* 1981, Phillips 1990) |
| 95. | $^3O + ^2C_2H(X^2\Sigma^+) \rightarrow ^1CO + ^2CH$ | -328 | 7.0e-11 | 0 | 0 | 2 | 0 | (Boullart *et al.* 1996, Devriendt *et al.* 1996) |
| 96. | $^3O + l\text{-}^2C_3H \rightarrow ^1CO + ^2C_2H$<br>$\rightarrow ^1C_3O + ^2H$ | -535<br>-442 | 5.0e-11<br>2.0e-11 | 0<br>0 | 0<br>0 | 2<br>2 | 0<br>0 | See text. |
| 97. | $^3O + c\text{-}^2C_3H \rightarrow ^1CO + ^2C_2H$<br>$\rightarrow ^1C_3O + ^2H$ | -522<br>-429 | 5.0e-11<br>2.0e-11 | 0<br>0 | 0<br>0 | 2<br>2 | 0<br>0 | See text. |
| 98. | $^3O + ^2C_4H \rightarrow ^1CO + ^2C_3H$<br>$\rightarrow ^1C_4O + ^2H$ | -447<br>-250 | 7.0e-11<br>0 | 0<br>0 | 0<br>0 | 2<br>2 | 0<br>0 | See text. |
| 99. | $^3O + ^2C_5H \rightarrow ^1CO + ^2C_4H$<br>$\rightarrow ^1C_5O + ^2H$ | -475<br>-321 | 6.0e-11<br>1.0e-11 | 0<br>0 | 0<br>0 | 2<br>2 | 0<br>0 | See text. |
| 100. | $^3O + ^2C_6H \rightarrow ^1CO + ^2C_5H$ | -440 | 7.0e-11 | 0 | 0 | 2 | 0 | See text. |
| 101. | $^3O + ^2C_7H \rightarrow ^1CO + ^2C_6H$<br>$\rightarrow ^1C_7O + ^2H$ | | 6.0e-11<br>1.0e-11 | 0<br>0 | 0<br>0 | 2<br>2 | 0<br>0 | See text. |
| 102. | $^3O + ^2C_8H \rightarrow ^1CO + ^2C_7H$ | | 7.0e-11 | 0 | 0 | 2 | 0 | See text. |
| 103. | $^3O + ^2C_9H \rightarrow ^1CO + ^2C_8H$<br>$\rightarrow ^1C_9O + ^2H$ | | 6.0e-11<br>1.0e-11 | 0<br>0 | 0<br>0 | 2<br>2 | 0<br>0 | See text. |
| 104. | $^3O + ^2C_{10}H \rightarrow ^1CO + ^2C_9H$ | | 7.0e-11 | 0 | 0 | 2 | 0 | See text. |
| | | | | | | | | |
| 105. | $^3O + ^1C_4H_2 \rightarrow ^1CO + t\text{-}^3C_3H_2$ | | 1.31e-11 | 0 | 678 | 1.6 | 100 | (Mitchell *et al.* 1986) |
| 106. | $^3O + ^1C_6H_2 \rightarrow ^1CO + ^3C_5H_2$ | | 3.0e-11 | -0.5 | 0 | 2 | 100 | See text. |
| 107. | $^3O + ^1C_8H_2 \rightarrow ^1CO + ^3C_7H_2$ | | 3.0e-11 | -0.5 | 0 | 2 | 100 | See text. |
| 108. | $^3O + ^1C_{5,7,9}H_2 \rightarrow ^1C_{4,6,8}H_2 + ^1CO$ | -699 | 2.0e-10 | 0 | 0 | 3 | 0 | See text. |
| | | | | | | | | |
| 109. | $^3O + CN(X^2\Sigma^+) \rightarrow ^1CO + N$ | -322 | 7.0e-11 | 0.17 | 0 | 2 | 0 | (Andersson et al. 2003, Titarchuk & Halpern 1995, |
| | | | | | | | | Schmatjko & Wolfrum 1977) |
| 110. | $^3O + C_2N(X^2\Pi) \rightarrow {}^1CO + {}^2CN$ | -614 | 7.0e-11 | 0.17 | 0 | 3 | 0 | See text. |
| 111. | $^3O + C_3N(X^2\Sigma^+) \rightarrow {}^1CO + {}^2C_2N$ | -361 | 7.0e-11 | 0.17 | 0 | 3 | 0 | See text. |
| 112. | $^3O + C_4N(X^2\Pi) \rightarrow {}^1CO + {}^2C_3N$ | -516 | 7.0e-11 | 0.17 | 0 | 3 | 0 | See text. |
| 113. | $^3O + C_5N(X^2\Sigma^+) \rightarrow {}^1CO + {}^2C_4N$ | -474 | 7.0e-11 | 0.17 | 0 | 3 | 0 | See text. |
| 114. | $^3O + C_6N(X^2\Pi) \rightarrow {}^1CO + {}^2C_5N$ | -438 | 7.0e-11 | 0.17 | 0 | 3 | 0 | See text. |
| 115. | $^3O + C_7N(X^2\Sigma^+) \rightarrow {}^1CO + {}^2C_6N$ | | 7.0e-11 | 0.17 | 0 | 3 | 0 | See text. |
| 116. | $^3O + C_8N(X^2\Pi) \rightarrow {}^1CO + {}^2C_7N$ | | 7.0e-11 | 0.17 | 0 | 3 | 0 | See text. |
| 117. | $^3O + C_9N(X^2\Sigma^+) \rightarrow {}^1CO + {}^2C_8N$ | | 7.0e-11 | 0.17 | 0 | 3 | 0 | See text. |
| 118. | $^3O + C_{10}N(X^2\Pi) \rightarrow {}^1CO + {}^2C_9N$ | | 7.0e-11 | 0.17 | 0 | 3 | 0 | See text. |
| | | | | | | | | |
| 119. | $^3C + OH(X^2\Pi) \rightarrow {}^1CO + {}^2H$ | | 1.15e10 | -0.34 | -0.1 | 2 | 7 | (Zanchet *et al.* 2009) |
| 120. | $^3O + OH(X^2\Pi) \rightarrow {}^3O_2 + {}^2H$ | | 2.0e-11 | 0 | 0 | 2 | 7 | 10K only, (Lique *et al.* 2009, Quéméner *et al.* 2009) |
| 121. | $^4N + OH(X^2\Pi) \rightarrow {}^2NO + {}^2H$ | | 6.6e-11 | 0.26 | 0 | 1.4 | 7 | (Daranlot *et al.* 2011) |
| 122. | $^4N + NO(X^2\Pi) \rightarrow {}^1N_2 + {}^3O$ | | 3.8e-11 | -0.27 | 23 | 1.4 | 7 | (Bergeat *et al.* 2009) |
| | | | | | | | | |
| | | | | | | | | |
| | **ionic changes:** | | | | | | | |
| 123. | $HCNH^+ + C_nH^- \rightarrow HCN + C_nH + H$<br>$\rightarrow HNC + C_nH + H$ | -252<br>-197 | 3.8e-8<br>3.8e-8 | -0.5<br>-0.5 | 0<br>0 | 2<br>2 | 0<br>0 | |
| 124. | $HCNH^+ + C_n^- \rightarrow HCN + C_n + H$<br>$\rightarrow HNC + C_n + H$ | | 3.8e-8<br>3.8e-8 | -0.5<br>-0.5 | 0<br>0 | 2<br>2 | 0<br>0 | |
| | **$C_{2n+1}O^+$ family:** | | | | | | | |
| 125. | $C_3O + H_3^+ \rightarrow HC_3O^+ + H_2$ | -458 | 1.0 | 3.42e-9 | 3.49 | 2 | 0 | Thermochemistry from (Hunter & Lias 1998) and calculations at the M06-2X/cc-pVTZ level using Gaussian09 (this work). Ionpol1. |
| 126. | $C_5O + H_3^+ \rightarrow HC_5O^+ + H_2$ | | 1.0 | 2.40e-9 | 4.02 | 2 | 0 | Ionpol1 |
| 127. | $C_7O + H_3^+ \rightarrow HC_7O^+ + H_2$ | | 1.0 | 2.40e-9 | 4.02 | 2 | 0 | Ionpol1, same as $C_5O + H_3^+$. |
| 128. | $C_9O + H_3^+ \rightarrow HC_9O^+ + H_2$ | | 1.0 | 2.40e-9 | 4.02 | 2 | 0 | Ionpol1, same as $C_5O + H_3^+$. |
| 129. | $C_3O + HCO^+ \rightarrow HC_3O^+ + CO$ | -286 | 1.0 | 1.33e-9 | 3.49 | 2 | 0 | Thermochemistry from (Hunter & Lias 1998) and calculations at the M06-2X/cc-pVTZ level using |
| | | | | | | | | Gaussian09 (this work). Ionpol1 |
| 130. | $C_5O + HCO^+ \rightarrow HC_5O^+ + CO$ | | 1.0 | 1.68e-9 | 4.02 | 2 | 0 | Ionpol1 |
| 131. | $C_7O + HCO^+ \rightarrow HC_7O^+ + CO$ | | 1.0 | 1.68e-9 | 4.02 | 2 | 0 | Ionpol1, same as $C_5O + HCO^+$. |
| 132. | $C_9O + HCO^+ \rightarrow HC_9O^+ + CO$ | | 1.0 | 1.68e-9 | 4.02 | 2 | 0 | Ionpol1, same as $C_5O + HCO^+$. |
| 133. | $C_3O + HCNH^+ \rightarrow HC_3O^+ + HCN$<br>$\rightarrow HC_3O^+ + HNC$ | -167<br>-112 | 0.5<br>0.5 | 1.34e-9<br>1.34e-9 | 3.49<br>3.49 | 2<br>2 | 0<br>0 | Thermochemistry from (Hunter & Lias 1998) and calculations at the M06-2X/cc-pVTZ level using Gaussian09 (this work). Ionpol1 |
| 134. | $C_5O + HCNH^+ \rightarrow HC_5O^+ + HCN$<br>$\rightarrow HC_5O^+ + HNC$ | | 0.5<br>0.5 | 1.71e-9<br>1.71e-9 | 4.02<br>4.02 | 2<br>2 | 0<br>0 | Ionpol1 |
| 135. | $C_7O + HCNH^+ \rightarrow HC_7O^+ + HCN$<br>$\rightarrow HC_7O^+ + HNC$ | | 0.5<br>0.5 | 1.71e-9<br>1.71e-9 | 4.02<br>4.02 | 2<br>2 | 0<br>0 | Ionpol1, same as $C_5O + HCNH^+$. |
| 136. | $C_9O + HCNH^+ \rightarrow HC_9O^+ + HCN$<br>$\rightarrow HC_9O^+ + HNC$ | | 0.5<br>0.5 | 1.71e-9<br>1.71e-9 | 4.02<br>4.02 | 2<br>2 | 0<br>0 | Ionpol1, same as $C_5O + HCNH^+$. |
| 137. | $HC_{3,5,7,9}O^+ + e^- \rightarrow H + C_{3,5,7,9}O$<br>$\rightarrow C_{2,4,6,8}H + CO$ | | 2.0e-7<br>2.0e-7 | -0.5<br>-0.5 | 0<br>0 | 3<br>3 | 0<br>0 | By comparison with similar reactions (Florescu-Mitchell & Mitchell 2006) |
| | **$C_{2n+1}N^+$ family:** | | | | | | | |
| 138. | $HCN + C^+ \rightarrow H + CNC^+$<br>$\rightarrow H + C_2N^+$ | -100<br>+6 | 1.0<br>0 | 1.28e-9 | 6.6 | 2 | 0 | Ionpol1, (Clary *et al.* 1990, Clary *et al.* 1985, Anicich *et al.* 1986) |
| 139. | $HNC + C^+ \rightarrow H + CNC^+$<br>$\rightarrow H + C_2N^+$ | -155<br>-49 | 0.5<br>0.5 | 1.34e-9<br>1.34e-9 | 6.5<br>6.5 | 2<br>2 | 0<br>0 | Ionpol1 |
| 140. | $HC_3N + C^+ \rightarrow H + C_4N^+$<br>$\rightarrow C_3H^+ + CN$<br>$\rightarrow HCN + C_3^+$ | ≈-120<br>≈-130<br>≈+60 | 0.20<br>0.80<br>0 | 1.82e-9<br>1.82e-9 | 5.44 | 2<br>2 | 0<br>0 | Ionpol1, (Anicich 2003). There are likely to be various $C_4N^+$ isomers ($C_3NC^+$). |
| 141. | $HC_5N + C^+ \rightarrow H + C_6N^+$<br>$\rightarrow C_5H^+ + CN$<br>$\rightarrow HCN + C_5^+$ | ≈-120<br>≈-130<br>≈+60 | 0.20<br>0.80<br>0 | 2.26e-9<br>2.26e-9 | 5.10 | 2<br>2 | 0<br>0 | Ionpol1, branching ratio from $HC_3N + C^+$ |
| 142. | $HC_7N + C^+ \rightarrow H + C_8N^+$<br>$\rightarrow C_7H^+ + CN$<br>$\rightarrow HCN + C_7^+$ | ≈-120<br>≈-130<br>≈+60 | 0.20<br>0.80<br>0 | 2.82e-9<br>2.82e-9 | 4.52 | 2<br>2 | 0<br>0 | Ionpol1, branching ratio from $HC_3N + C^+$ |
| 143. | $HC_9N + C^+ \rightarrow H + C_{10}N^+$<br>$\rightarrow C_9H^+ + CN$<br>$\rightarrow HCN + C_9^+$ | ≈-120<br>≈-130<br>≈+60 | 0.20<br>0.80<br>0 | 2.82e-9<br>2.82e-9 | 4.52 | 2<br>2 | 0<br>0 | Ionpol1, same as $HC_7N + C^+$. |
| | **$C_{2,4,6,8,10}N$ ionic reactions:** | | | | | | | |
| 144. | $H_2 + C_2N^+ \rightarrow HCNH^+ + C$ | -60 | 9.0e-10 | 0 | 0 | 1.6 | 0 | (Knight *et al.* 1988) |
| 145. | $H_2 + CNC^+ \rightarrow HCNH^+ + C$ | +46 | 0 | 0 | 0 | 0 | 0 | Endothermic (Knight *et al.* 1988) |
| 146. | $H_2 + C_4N^+ \rightarrow HCN + C_3H^+$<br>$\rightarrow HNC + C_3H^+$<br>$\rightarrow HCNH^+ + C_3$ | ≈ -250<br>≈ -200<br>≈ -200 | 3.0e-10<br>3.0e-10<br>3.0e-10 | 0<br>0<br>0 | 0<br>0<br>0 | 2<br>2<br>2 | 0<br>0<br>0 | Same as $H_2 + C_2N^+$. |
| 147. | $H_2 + C_6N^+ \rightarrow HCN + C_5H^+$<br>$\rightarrow HNC + C_5H^+$<br>$\rightarrow HCNH^+ + C_5$ | ≈ -250<br>≈ -200<br>≈ -200 | 3.0e-10<br>3.0e-10<br>3.0e-10 | 0<br>0<br>0 | 0<br>0<br>0 | 2<br>2<br>2 | 0<br>0<br>0 | Same as $H_2 + C_2N^+$. |
| 148. | $H_2 + C_8N^+ \rightarrow HCN + C_7H^+$<br>$\rightarrow HNC + C_7H^+$<br>$\rightarrow HCNH^+ + C_7$ | ≈ -250<br>≈ -200<br>≈ -200 | 3.0e-10<br>3.0e-10<br>3.0e-10 | 0<br>0<br>0 | 0<br>0<br>0 | 2<br>2<br>2 | 0<br>0<br>0 | Same as $H_2 + C_2N^+$. |
| 149. | $H_2 + C_{10}N^+ \rightarrow HCN + C_9H^+$<br>$\rightarrow HNC + C_9H^+$<br>$\rightarrow HCNH^+ + C_9$ | ≈ -250<br>≈ -200<br>≈ -200 | 3.0e-10<br>3.0e-10<br>3.0e-10 | 0<br>0<br>0 | 0<br>0<br>0 | 2<br>2<br>2 | 0<br>0<br>0 | Same as $H_2 + C_2N^+$. |
| | | | | | | | | |
| 150. | $CCN + C^+ \rightarrow CCN^+ + C$<br>$\rightarrow CNC^+ + C$<br>$\rightarrow CN + C_2^+$<br>$\rightarrow CN^+ + C_2$ | -49<br>-155<br>-81<br>-7 | 0.3<br>0.4<br>0.3<br>0 | 1.30e-9<br>1.30e-9<br>1.30e-9 | 0.62<br>0.62<br>0.62 | 2<br>2<br>2 | 10<br>10<br>10 | (Mebel & Kaiser 2002, Harland & McIntosh 1985).<br>10-30K: Ionpol1, 40-300K: Ionpol2 |
| 151. | $C_4N + C^+ \rightarrow C_3 + CNC^+$<br>$\rightarrow C_3 + CCN^+$<br>$\rightarrow C_4N^+ + C$ | -199<br>-93<br>-138 | 0.6<br>0.1<br>0.3 | 2.05e-9<br>2.05e-9<br>2.05e-9 | 0<br>0<br>0 | 2<br>2<br>2 | 10<br>10<br>10 | Ionpol2 |
| 152. | $C_6N + C^+ \rightarrow C_5 + CNC^+$<br>$\rightarrow C_5 + CCN^+$<br>$\rightarrow C_6N^+ + C$ | | 0.6<br>0.1<br>0.3 | 2.71e-9<br>2.71e-9<br>2.71e-9 | 0.25<br>0.25<br>0.25 | 2<br>2<br>2 | 10<br>10<br>10 | Ionpol2 |
| 153. | $C_8N + C^+ \rightarrow C_7 + CNC^+$<br>$\rightarrow C_7 + CCN^+$<br>$\rightarrow C_8N^+ + C$ | | 0.6<br>0.1<br>0.3 | 3.35e-9<br>3.35e-9<br>3.35e-9 | 0.43<br>0.43<br>0.43 | 2<br>2<br>2 | 10<br>10<br>10 | 10-20K: Ionpol1<br>30-300K: Ionpol2 |
| 154. | $C_{10}N + C^+ \rightarrow C_9 + CNC^+$<br>$\rightarrow C_9 + CCN^+$<br>$\rightarrow C_{10}N^+ + C$ | | 0.6<br>0.1<br>0.3 | 3.35e-9<br>3.35e-9<br>3.35e-9 | 0.43<br>0.43<br>0.43 | 2<br>2<br>2 | 10<br>10<br>10 | 10-20K: Ionpol1<br>30-300K: Ionpol2 |
| | | | | | | | | |

| | | | | | | | | | |
|---|---|---|---|---|---|---|---|---|---|
| 155. | $C_2N + H_3^+$ | $\rightarrow HC_2N^+ + H_2$ | -270 | 1.0 | 2.90e-9 | 0.62 | 2 | 0 | Thermochemistry from (Hunter & Lias 1998) and calculations at the M06-2X/cc-pVTZ level using Gaussian09 (this work). 10-20K: Ionpol1, 30-300K: Ionpol2 |
| 156. | $C_4N + H_3^+$ | $\rightarrow HC_4N^+ + H_2$ | -367 | 1.0 | 3.85e-9 | 0.10 | 2 | 0 | Thermochemistry from (Hunter & Lias 1998) and calculations at the M06-2X/cc-pVTZ level using Gaussian09 (this work). Ionpol2 |
| 157. | $C_6N + H_3^+$ | $\rightarrow HC_6N^+ + H_2$ | | 1.0 | 5.16e-9 | 0.25 | 2 | 0 | Ionpol2 |
| 158. | $C_8N + H_3^+$ | $\rightarrow HC_8N^+ + H_2$ | | 1.0 | 7.31e-9 | 0.38 | 2 | 0 | 10K: Ionpol1, 20-300K: Ionpol2 |
| 159. | $C_{10}N + H_3^+$ | $\rightarrow HC_{10}N^+ + H_2$ | | 1.0 | 7.31e-9 | 0.38 | 3 | 0 | 10K: Ionpol1, 20-300K: Ionpol2 |
| 160. | $C_2N + HCO^+$ | $\rightarrow HC_2N^+ + CO$ | -98 | 1.0 | 1.19e-9 | 0.62 | 2 | 0 | Thermochemistry from (Hunter & Lias 1998) and calculations at the M06-2X/cc-pVTZ level using Gaussian09 (this work). 10-20K: Ionpol1, 30-300K: Ionpol2 |
| 161. | $C_4N + HCO^+$ | $\rightarrow HC_4N^+ + CO$ | -195 | 1.0 | 1.47e-9 | 0.10 | 2 | 0 | Thermochemistry from (Hunter & Lias 1998) and calculations at the M06-2X/cc-pVTZ level using Gaussian09 (this work). Ionpol2 |
| 162. | $C_6N + HCO^+$ | $\rightarrow HC_6N^+ + CO$ | | 1.0 | 1.89e-9 | 0.25 | 2 | 0 | Ionpol2 |
| 163. | $C_8N + HCO^+$ | $\rightarrow HC_8N^+ + CO$ | | 1.0 | 2.61e-9 | 0.38 | 2 | 0 | 10K: Ionpol1, 20-300K: Ionpol2 |
| 164. | $C_{10}N + HCO^+$ | $\rightarrow HC_{10}N^+ + CO$ | | 1.0 | 2.61e-9 | 0.38 | 2 | 0 | Idem as $C_8N + HCO^+$. 10K: Ionpol1, 20-300K: Ionpol2 |
| 165. | $C_2N + HCNH^+$ | $\rightarrow HC_2N^+ + HCN$<br>$\rightarrow HC_2N^+ + HNC$ | +30<br>+89 | 0<br>0 | 0<br>0 | 0<br>0 | 0<br>0 | 0<br>0 | Thermochemistry from (Hunter & Lias 1998) and calculations at the M06-2X/cc-pVTZ level using Gaussian09 (this work). |
| 166. | $C_4N + HCNH^+$ | $\rightarrow HC_4N^+ + HCN$<br>$\rightarrow HC_4N^+ + HNC$ | -76<br>-21 | 0.50<br>0.50 | 1.48e-9<br>1.48e-9 | 0.10<br>0.10 | 2<br>2 | 0<br>0 | Thermochemistry from (Hunter & Lias 1998) and calculations at the M06-2X/cc-pVTZ level using Gaussian09 (this work). Ionpol2 |
| 167. | $C_6N + HCNH^+$ | $\rightarrow HC_6N^+ + HCN$<br>$\rightarrow HC_6N^+ + HNC$ | | 0.50<br>0.50 | 1.91e-9<br>1.91e-9 | 0.25<br>0.25 | 2<br>2 | 0<br>0 | Ionpol2 |
| 168. | $C_8N + HCNH^+$ | $\rightarrow HC_8N^+ + HCN$<br>$\rightarrow HC_8N^+ + HNC$ | | 0.50<br>0.50 | 2.64e-9<br>2.64e-9 | 0.38<br>0.38 | 2<br>2 | 0<br>0 | 10K: Ionpol1<br>20-300K: Ionpol2 |
| 169. | $C_{10}N + HCNH^+$ | $\rightarrow HC_{10}N^+ + HCN$<br>$\rightarrow HC_{10}N^+ + HNC$ | | 0.50<br>0.50 | 2.64e-9<br>2.64e-9 | 0.38<br>0.38 | 2<br>2 | 0<br>0 | Idem as $C_8N + HCNH^+$ |
| | | | | | | | | | |
| 170. | $H_2 + HC_2N^+$ | $\rightarrow H_2C_2N^+ + H$ | -88 | 1.0 | 1.51e-9 | 0 | 2 | 0 | Thermochemistry from (Scott et al. 1999, Holmes et al. |

| | | | | | | | | |
|---|---|---|---|---|---|---|---|---|
| | | | | | | | | 1993), Ionpol2 |
| 171. | $H_2 + HC_4N^+ \rightarrow H_2C_4N^+ + H$ | | 1.0 | 1.49e-9 | 0 | 2 | 0 | Ionpol2 |
| 172. | $H_2 + HC_6N^+ \rightarrow H_2C_6N^+ + H$ | | 1.0 | 1.49e-9 | 0 | 2 | 0 | Ionpol2 |
| 173. | $H_2 + HC_8N^+ \rightarrow H_2C_8N^+ + H$ | | 1.0 | 1.49e-9 | 0 | 2 | 0 | Ionpol2 |
| 174. | $H_2 + HC_{10}N^+ \rightarrow H_2C_{10}N^+ + H$ | | 1.0 | 1.49e-9 | 0 | 2 | 0 | Ionpol2 |
| 175. | $H_2 + H_2C_{2,4,6,8,10}N^+ \rightarrow H_3C_{2,4,6,8,10}N^+ + H$ | ≈ +200 | 0 | 0 | 0 | 0 | 0 | Thermochemistry calculated at the M06-2X/cc-pVTZ level using Gaussian09 (this work). |
| | | | | | | | | |
| 176. | $CNC^+ + e^- \rightarrow C + CN$<br>$\rightarrow N + C_2$ | -468<br>-309 | 3.8e-7<br>2.0e-8 | -0.6<br>-0.6 | 0<br>0 | 3<br>3 | 0<br>0 | KIDA datasheet |
| 177. | $C_2N^+ + e^- \rightarrow C + CN$<br>$\rightarrow N + C_2$ | -574<br>-415 | 2.0e-8<br>3.8e-7 | -0.6<br>-0.6 | 0<br>0 | 3<br>3 | 0<br>0 | By comparison with similar reactions (Mitchell et al. 1986). Branching ratios from (Plessis *et al.* 2012) . |
| 178. | $C_{4,6,8,10}N^+ + e^- \rightarrow CN + C_{3,5,7,9}$ | | 4.0e-7 | -0.6 | 0 | 3 | 0 | By comparison with similar reactions (Mitchell et al. 1986). Branching ratios from (Plessis *et al.* 2012). |
| 179. | $HC_2N^+ + e^- \rightarrow CH + CN$<br>$\rightarrow H + C_2N$<br>$\rightarrow C + HCN$<br>$\rightarrow C + HNC$ | -505<br>-629<br>-684<br>-629 | 1e-7<br>1e-7 | -0.5<br>-0.5 | 0<br>0 | 4<br>4 | 0<br>0 | By comparison with similar reactions (Mitchell et al. 1986). Branching ratios from (Plessis *et al.* 2012). |
| 180. | $HC_{4,6,8,10}N^+ + e^- \rightarrow l\text{-}C_{3,5,6,9}H + CN$<br>$\rightarrow H + C_{4,6,8,10}N$<br>$\rightarrow C_{3,5,6,9} + HCN$ | ≈ -400<br>≈-500<br>≈-600 | 1e-7<br>1e-7 | -0.5<br>-0.5 | 0<br>0 | 4<br>4 | 0<br>0 | By comparison with similar reactions (Mitchell et al. 1986). Branching ratios from (Plessis *et al.* 2012) . |
| 181. | $H_2C_2N^+ + e^- \rightarrow CH + HCN$<br>$\rightarrow CH_2 + CN$<br>$\rightarrow H + HCCN$ | -526<br>-438<br>-568 | 1e-7<br>1e-7<br>0 | -0.5<br>-0.5<br>0 | 0<br>0<br>0 | 4<br>4<br>0 | 0<br>0<br>0 | By comparison with similar reactions (Mitchell et al. 1986). Branching ratios from (Plessis *et al.* 2012).<br>HCCN not present in KIDA |
| 182. | $H_2C_{4,6,8,10}N^+ + e^- \rightarrow CH + HC_{3,5,7,9}N$<br>$\rightarrow C_2H_2 + C_{2,4,6}N$<br>$\rightarrow l\text{-}C_{3,5,7,9}H_2 + CN$ | | 1e-7<br>1e-7<br>1e-7 | -0.5<br>-0.5<br>-0.5 | 0<br>0<br>0 | 4<br>4<br>4 | 0<br>0<br>0 | By comparison with similar reactions (Mitchell et al. 1986). Branching ratios from (Plessis *et al.* 2012). |

Andersson S., Markovic N., Nyman G., 2003, J. Phys. Chem A., 107**,** 5439

Anicich V.G., 2003, JPL Publication 2003, 03-19 NASA.,

Anicich V.G., Huntress W.T., McEwan M.J., 1986, J. Phys. Chem., 90**,** 2446

Becker K.H., Donner B., Dinis C.M.F., Geiger H., Schmidt F., Wiesen P., 2000, Z. Phys. Chem., 214**,** 503

Bergeat A., Hickson K.M., Daugey N., Caubet P., Costes M., 2009, Phys. Chem. Chem. Phys., 11**,** 8149

Boggio-Pasqua M., Voronin A.I., Halvick P., Rayez J.C., 2000, Phys. Chem. Chem. Phys., 2, 1693
Boullart W., Devriendt K., Borms R., Peteers J., 1996, J. Phys. Chem., 100, 998
Brownsword R.A., Gatenby S.D., Herbert L.B., Smith I.W.M., Stewart D.W.A., Symonds A.C., 1996, J. Chem. Soc. Faraday Trans., 7234
Chastaing D., Le Picard S.D., Sims I.R., Smith I.W.M., 2001, A&A, 365, 241
Clary D.C., Dateo C.E., Smith D., 1990, Chem. Phys. Lett., 167, 1
Clary D.C., Smith D., Adams N.G., 1985, Chem. Phys. Lett., 119, 320
Costes M., Halvick P., Hickson K.M., Daugey N., Naulin C., 2009, ApJ, 703, 1179
Daranlot J., Hincelin U., Bergeat A., Costes M., Loison J.-C., Wakelam V., Hickson K.M., 2012, PNAS, 109, 10233
Daranlot J., Jorfi M., Xie C., Bergeat A., Costes M., Caubet P., Xie D., Guo H., Honvault P., Hickson K.M., 2011, Science, 334, 1538
Devriendt K., Van Look H., Ceursters B., Peeters J., 1996, Chem. Phys. Lett., 261, 450
Diaz-Tendero S., Sanchez G., Alcami M., Martin F., Hervieux P.A., Chabot M., Martinet G., Désesquelles P., Mezdari F., Wohrer-Béroff K., Negra S.D., Hamrita H., LePadellec A., Montagnon L., 2006, Int. J. Mass Spectrom., 252, 126
Florescu-Mitchell A.I., Mitchell J.B.A., 2006, Phys. Rep., 430, 277
Gannon K.L., Glowacki D.R., Blitz M.A., Hughes K.J., Pilling M.J., Seakins P.W., 2007, J. Phys. Chem A., 111, 6679
Gu X., Guo Y., Mebel A.M., Kaiser R.I., 2007, Chem. Phys. Lett., 449, 44
Harland P.W., McIntosh B.J., 1985, Int. J. Mass Spectrom. Ion Proc., 67, 29
Holmes J.L., Lossing F.P., Mayer P.M., 1993, Chem. Phys. Lett., 212, 134
Hunter E.P.L., Lias S.G., 1998, J. Phys. Chem. Ref. Data, 27, 413
Knight J.S., Petrie S.A.H., Freeman C.G., McEwan M.J., McLean A.D., DeFrees D.J., 1988, JACS, 110, 5286
Li H.Y., Cheng W.C., Liu Y.L., Sun B.J., Huang C.Y., Chen K.T., Tang M.S., Kaiser R.I., Chang A.H.H., 2006, J. Chem. Phys., 124, 044307
Lique F., Jorfi M., Honvault P., Halvick P., Lin S.Y., Guo H., Xie D.Q., Dagdigian P.J., Klos J., Alexander M.H., 2009, J. Chem. Phys., 131, 221104
Mebel A.M., Kaiser R.I., 2002, ApJ, 564, 787
Messing I., Carrington T., Filseth S.V., Sadowski C.M., 1980, Chem. Phys. Lett., 74, 56
Messing I., Filseth S.V., Sadowski C.M., Carrington T., 1981, J. Chem. Phys., 74, 3874
Mitchell M.B., Nava D.F., Stief L.J., 1986, J. Chem. Phys., 85, 3300
Nelson H.H., Helvajian H., Pasternack L., McDonald J.R., 1982, Chem. Phys., 73, 431
Petrie S., Osamura Y., 2004, J. Phys. Chem. A, 108, 3623
Phillips L.F., 1990, Chem. Phys. Lett., 165, 545
Plessis S., Carrasco N., Dobrijevic M., Pernot P., 2012, Icarus, 219, 254
Quéméner G., Balakrishnan N., Kendrick B.K., 2009, Phys. Rev. A, 79, 022703
Schmatjko K.J., Wolfrum J., 1977, Int. Symp. Comb., 16, 819

Scott G.B.I., Fairley D.A., Freeman C.G., McEwan M.J., Anicich V.G., 1999, J. Phys. Chem. A, 103**,** 1073
Silva R., Gichuhi W.K., Kislov V.V., Landera A., Mebel A.M., Suits A.G., 2009, J. Phys. Chem. A, 113**,** 11182
Sims I.R., Queffelec J.-L., Travers D., Rowe B.R., Herbert L.B., Karthäuser J., Smith I.W.M., 1993, Chem. Phys. Lett., 211**,** 461
Titarchuk T.A., Halpern J.B., 1995, Chem. Phys. Lett., 232**,** 192
Van Orden A., Saykally R.J., 1998, Chem. Rev., 98**,** 2313
Wakelam V., Loison J.C., Herbst E., Talbi D., Quan D., Caralp F., 2009, A&A, 495**,** 513
Wang J., Ding Y.-h., Sun C.-c., 2006, ChemPhysChem, 7**,** 710
Zanchet A., Bussery-Honvault B., Jorfi M., Honvault P., 2009, Phys. Chem. Chem. Phys., 11**,** 6182